\documentclass[twocolumn]{aastex631}

\received{October 2, 2020}
\accepted{September 9, 2021}
\submitjournal{The Planetary Science Journal}

\graphicspath{{./}{figures/}}

\begin{document}

%%%%%%%%%%%%%%%%%%%%%%%%%%%%%%%%%%%%%%%%%%%%%%%%%%%%%%%%%%%%%%%%%%%%%%

\title{Thermophysical and Compositional Analyses of Dunes at Hargraves Crater, Mars}

%%%% Authors
\correspondingauthor{A. Emran}
\email{alemran@uark.edu}

\author[0000-0002-4420-0595]{A. Emran}
\affiliation{AR Center for Space and Planetary Sciences, University of Arkansas, Fayetteville, AR 72701, USA}
\affiliation{Department of Geosciences, Auburn University, Auburn, AL 36849, USA}

\author[0000-0002-3645-3672]{ L. J. Marzen}
\affiliation{Department of Geosciences, Auburn University, Auburn, AL 36849, USA}

\author[0000-0001-9939-7543]{D. T. King Jr.}
\affiliation{Department of Geosciences, Auburn University, Auburn, AL 36849, USA}

\author[0000-0002-1111-587X]{ V. F. Chevrier}
\affiliation{AR Center for Space and Planetary Sciences, University of Arkansas, Fayetteville, AR 72701, USA}

\begin{abstract}

We analyze thermal emission spectra using the 2001 Mars Odyssey Thermal Emission Imaging System (THEMIS) and the Mars Global Surveyor (MGS) Thermal Emission Spectrometer (TES) to characterize grain-size and mineralogical composition of dunes at Hargraves crater, Mars. Thermal inertia and bulk composition of the dunes were compared to inferred provenances from the thermal infrared response of surface constituent materials. We use a Markov Chain Monte Carlo (MCMC) technique to estimate the bulk amount of mineralogy contributed by each inferred provenance to the dune field composition. An average thermal inertia value of 238$\mathrm{\pm}$17 Jm${}^{-2}$K${}^{-1}$s${}^{-0.5}$ was found for the dunes corresponding to a surface composed of an average effective grain-size of $\mathrm{\sim}$391$\mathrm{\pm}$172 $\mu$m. This effective particle size suggests the presence of mostly medium sand-sized materials mixed with fine and coarse grain sands. The dunes are likely comprised of a weakly indurated surface mixed with unconsolidated materials. Compositional analysis specifies that the dunes are comprised of a mixture of feldspar, olivine, pyroxene, and relatively low bulk-silica content. Dune materials were likely derived from physical weathering, especially eolian erosion, predominantly from the crater ejecta unit at the crater, mixed with a small amount from the crater floor and crater rim and wall lithologies- indicating the dune materials were likely sourced locally.

\end{abstract}
\keywords{Solar system planets (1260) --- Solar system astronomy (1529) --- Solar system (1528) --- Inner planets (797) --- Mars (1007)}
%\TheUnifiedAstronomyThesaurus{Solar system planets (1260) --- Solar system astronomy (1529) --- Solar system (1528) --- Inner planets (797) --- Mars (1007) --- Solar system (1528)}

%%%%%%%%%%%%%%%%%%%%%%%%%%%%%%%%%%%%%%%%%%%%%%%%%%%%%%%%%%%%%%%%%%%%%%%%
\section{Introduction} \label{sec:intro}
Despite a variety of environments involved, dune fields provide an important insight into eolian transport regimes (both past and present) of terrestrial planets \citep{greeley1987wind}. The surface of Mars is rife with eolian dune fields in a variety of locations e.g., the circum-north polar region \citep{edgett1991particle}. Dunes on the martian surface were first detected during the early 1970s \citep{sagan1972variable}. The advent of higher resolution orbital images further confirmed the presence of dunes during the late 1990s \citep{edgett2000new} and showed similarities with terrestrial dunes \citep{bandeira2010automated}. Recently, a global martian dune database called the Mars Global Digital Dune Database (MGD${}^{3}$) \citep{hayward2007amars, hayward2010mars, hayward2012mars} has been prepared by planetary scientists. The database contains spatial and morphological characteristics of identified dune fields of $\mathrm{>}$1 km${}^{2}$ \citep{hayward2007bmars, hayward2014mars}. Understanding of martian dune materials and eolian processes opens a window for constraining the surface-atmospheric interactions, weather, climate, and climatic evolution of the planet \citep{greeley2001aeolian, wilson2004latitude}. Over the past decades, studies were carried out to better understand martian surface processes and global scale atmospheric dynamics from analysis of various eolian processes and wind-related features \citep{ward1985global, fenton2005aeolian, hayward2009aeolian, silvestro2010dune, gardin2012dune, sefton2014constraints}.

Throughout its geologic history, the surface of Mars has been shaped by eolian processes. Thus, analyses of thermophysical, grain-size, and mineral compositional distribution within dunes are regarded as important aspects for helping to constrain the geology and climate of the planet \citep{greeley1987wind, greeley2001aeolian, charles2017comparison, fenton2019bmars}. Local and regional scale analysis of grain-size distributions is important in classifying sedimentary environments and understanding sediment transport dynamics. Particle grain-size provides important information about shear stress applied by the medium of transport to initiate and sustain particle movement \citep{abuodha2003grain}. Particle grain-size in dunes is actively influenced by nature of source materials, surface topography, transport medium, and the distance traveled from source to sink \citep{abuodha2003grain}. On Mars, grain-size distribution has a direct relationship with thermophysical characteristics of constituent materials \citep{kieffer1977thermal, haberle1991atmospheric}. Eolian dunes of Mars are likely comprised of particle-sizes of homogenous materials within a dune field and, therefore, grain-size determination from thermal inertia is considered a reliable method \citep{edwards2018thermophysical}.

The approximate bulk particle size of dunes on Mars have been determined successfully using thermal inertia from the Thermal Emission Imaging System (THEMIS) images \citep{fenton2006thermal, fergason2006high, fergason2012surface}. \cite{edwards2018thermophysical} estimated the average grain size for the Bagnold dune field to be $\mathrm{\sim}$250$\mathrm{\pm}$85 $\mu$m with a minimum of 148 $\mu$m and a maximum of 968 $\mu$m based on derived thermal inertia values. Using data from Mars Science Laboratory (MSL) Curiosity, \cite{ehlmann2017chemistry} reported presence of very fine to medium-sized ($\mathrm{\sim}$45-500 $\mu$m) sand particles in the active Bagnold dune field. THEMIS-derived thermal inertia of $\mathrm{\sim}$250 - 410 Jm${}^{-2}$K${}^{-1}$s${}^{-0.5}$ was found across the majority of the Gale crater landing site of MSL, which corresponds to dunes comprised of an indurated surface, likely mixed with unconsolidated materials \citep{fergason2012surface}. In addition, Mawrth Vallis has an average thermal inertia of 310 Jm${}^{-2}$K${}^{-1}$s${}^{-0.5}$, indicating${}^{\ }$a mixture of bedrock, bedform, indurated surface, and unconsolidated materials \citep{fergason2012surface}.

Analysis of thermal inertia and local dune morphology variation provides insight into near-surface composition and local to regional climate characteristics \citep{courville2016thermophysical}, for instance, the presence of subsurface ice or volatiles. Besides thermal inertia derived from THEMIS nighttime thermal infrared (TIR) measurements, higher resolution images compared to THEMIS, such as the Context Camera (CTX) \citep{malin2007context} and the High Resolution Imaging Science Experiment (HiRISE) \citep{mcewen2007mars}, can be used as a visual survey tool for interpretation of thermal inertia \citep{fergason2012surface} and analyzing local tonal variability of dune fields. Compositional analysis of dune materials is key to better understanding sediment source and transport history. Bulk composition of dunes can reveal the extent of similarities (and dissimilarities) between eolian sands and surrounding surface compositions, provenances (sources) of dune materials, transport behavior of eolian medium, eolian activity level, and alteration (if any) of eolian sands versus their source materials \citep{fenton2019amars, fenton2019bmars}.

\cite{fenton2019bmars} did a complimentary science publication to MGD${}^{3}$ which describes mineralogy and morphologic stability of 79 large dune fields from thermal infrared (TIR) spectra at nearly global scale (80.7$\mathrm{{}^\circ}$S - 41.7$\mathrm{{}^\circ}$N). \cite{gullikson2018mars} expanded MGD${}^{3}$ to include thermal inertia and compositional information of dune fields in the equatorial and south polar regions. However, the latest installments of MGD${}^{3}$ only included compositions for dune fields $\mathrm{\ge}$300 km${}^{2}$ and, therefore, compositional analysis for dune fields smaller than that threshold is warranted. Also, there is a need to compare dune field composition with surrounding materials to determine whether their potential provenance is at a local scale versus regional or global. The outlines of dune fields in MGD${}^{3}$ are not always accurate and overlooks the recognition of smaller dune forms \citep{emran2020semiautomated} likely due to the low-resolution THEMIS imagery used and the possibility of human error when manually outlining the dune fields. Thus, extracting information from an accurate dune outline is a challenge. 

With increasing image resolution, the accuracy of outlining dune fields also increases. Recently, \cite{emran2020semiautomated} accurately outlined an MGD${}^{3}$ dune field at Hargraves crater from CTX data using a semi-automated method to include small dune forms present within the crater that were overlooked by MGD${}^{3\ }$defined outline. Their research provides an excellent opportunity to study a small dune field (area of $\mathrm{<}$ 100 km${}^{2}$) and compare it with potential provenances. The crater is located in the Nili Fossae region of the northwest Isidis basin (Fig. 1). The region has received substantial attention for planetary exploration missions and has been extensively studied over the past decade (\cite{salvatore2018bulk}, and references therein). Jezero crater, the recently landed site of Mars 2020 Perseverance rover, is also located in this region. Hargraves crater is considered to be part of the Jezero watershed \citep{goudge2015assessing}, though relatively little attention has been paid to this crater in comparison to Jezero crater and Nili Fossae.

\begin{figure*}[ht!]
\plotone{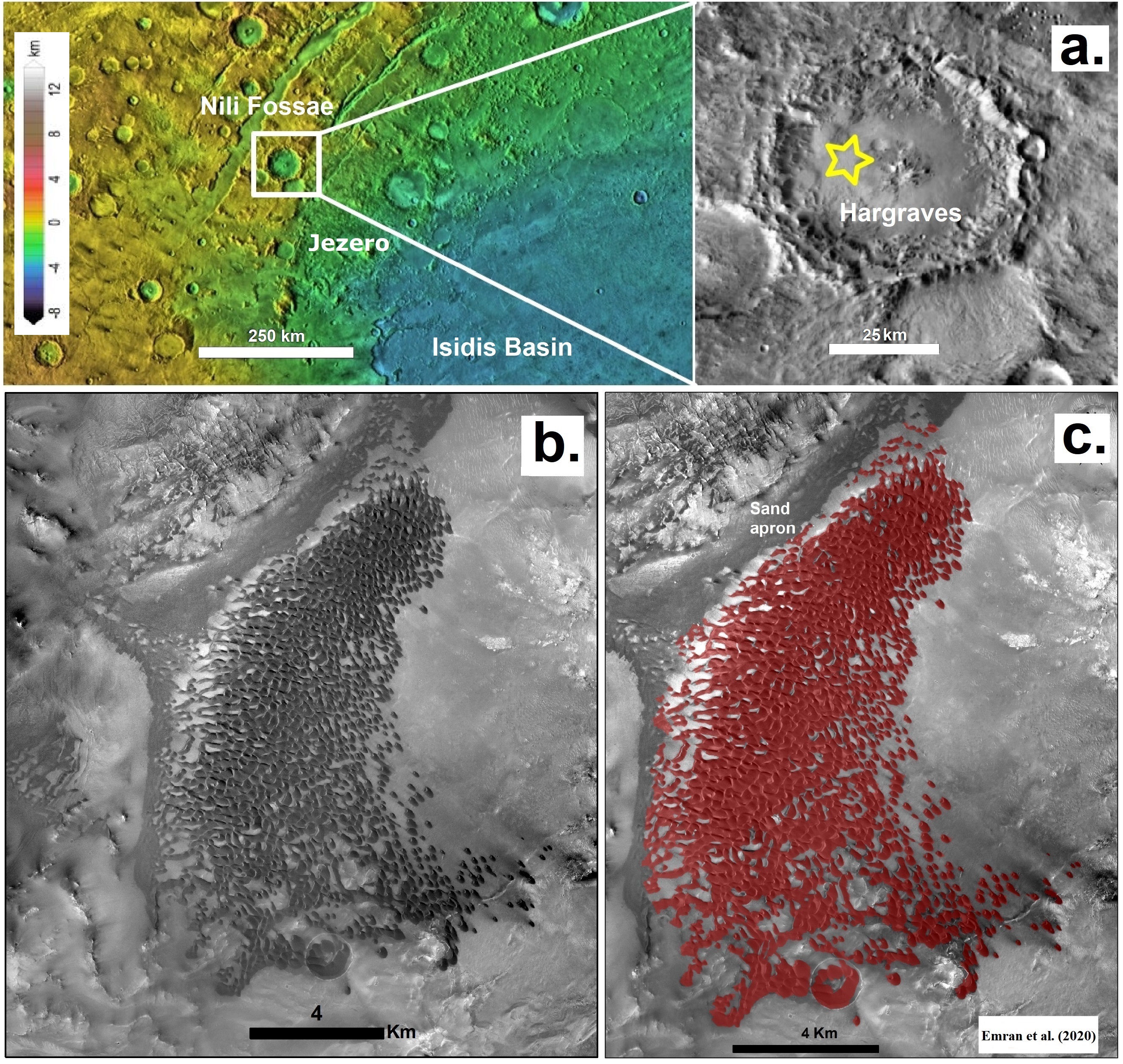}
\caption{a) Hargraves crater is located east of Nili Fossae and northwest of Isidis Basin, and is considered to be part of the Jezero watershed \citep{goudge2015assessing}. The background mosaic consists of MOLA colorized elevation data overlain onto a THEMIS image mosaic. The enlarged location (a) is shown in THEMIS daytime infrared images where the yellow star indicates the location of the dune field within the crater. b) The dune field in CTX image resolution. c) Dune polygon layer as derived by \cite{emran2019b}. The red polygons highlight the dunes within the crater. The figure is recreated after \cite{emran2020semiautomated}. North is up.  \label{fig:01}}
\end{figure*}

We determine thermophysical and mineralogical characteristics of the dunes within Hargraves. The dune field outline in \cite{emran2020semiautomated} was chosen as the boundary of dune materials in this study (the dune database is found in \cite{emran2019b}). We analyze the surface grain-size distribution and bulk compositional (mineral abundances) characteristics across the dune field. Active dunes preclude the possibility of cementation and vertical variation of constituent materials and, therefore, thermal inertia measurements can be used for grain-size analysis \citep{edwards2018thermophysical}. Grain size analysis derived from THEMIS thermal inertia is considered to be a reliable measurement in this study because the dune field within Hargraves is both potentially active \citep{emran2020semiautomated} and covers a large area, i.e., its spatial extent spans across numerous THEMIS pixels. 

We use CTX ($\mathrm{\sim}$6 m/pixel) and HiRISE (25-60 cm/pixel) images as a visual survey tool for analyzing tonal variability and presence of induration morphology i.e., degradational characteristics, such as erosional features (for example, mass wasting) or impact cratering as well as surface mantling within the dune field. We determine morphologic stability of the dune field that measures the degree of modification by non-eolian processes responsible for erosion and stabilization of the dunes \citep{fenton2019bmars}. Martian dunes are classified into six stability index (SI) categories based on an inferred scale from 1 to 6 constructed upon presence (or lack) of superposed non-eolian features (such as gullies, small pits, mass wasting features, etc.) and apparent level of degradation by non-eolian processes i.e., erosion \citep{banks2018patterns, fenton2019amars}. Assigning a stability index is a subjective procedure; dunes with no apparent stability features (i.e., an active dune field) are assigned a SI of 1 whereas SI values from 2 to 6 indicate increasing non-eolian modification \citep{fenton2019amars}. Refer to \cite{fentonhayward2010southern} for details of dune stability classes and their characteristics.

We then compare the bulk mineral assemblage of the dune field to surrounding geologic units inside the crater \citep{goudge2015assessing}, which have been inferred as potential source materials. Martian dune sand is believed to have not traveled far from their sources and reflect varying mineralogy of surrounding local surface terrain \citep{fenton2019bmars}. Thus, we hypothesize that the dunes at Hargraves are also locally sourced within the crater. Accordingly, we compare dune mineralogy to the surrounding geologic units at the crater using spectral data from Thermal Emission Spectrometer (TES) \citep{christensen2001mars} instrument and then subsequently modeled to determine the compositions.

%%%%%%%%%%%%%%%%%%%%%%%%%%%%%%%%%%%%%%%%%%%%%%%%%%%%%%%%%
\section{Hargraves crater} \label{sec:hargraves}
\begin{figure*}[ht!]
\plotone{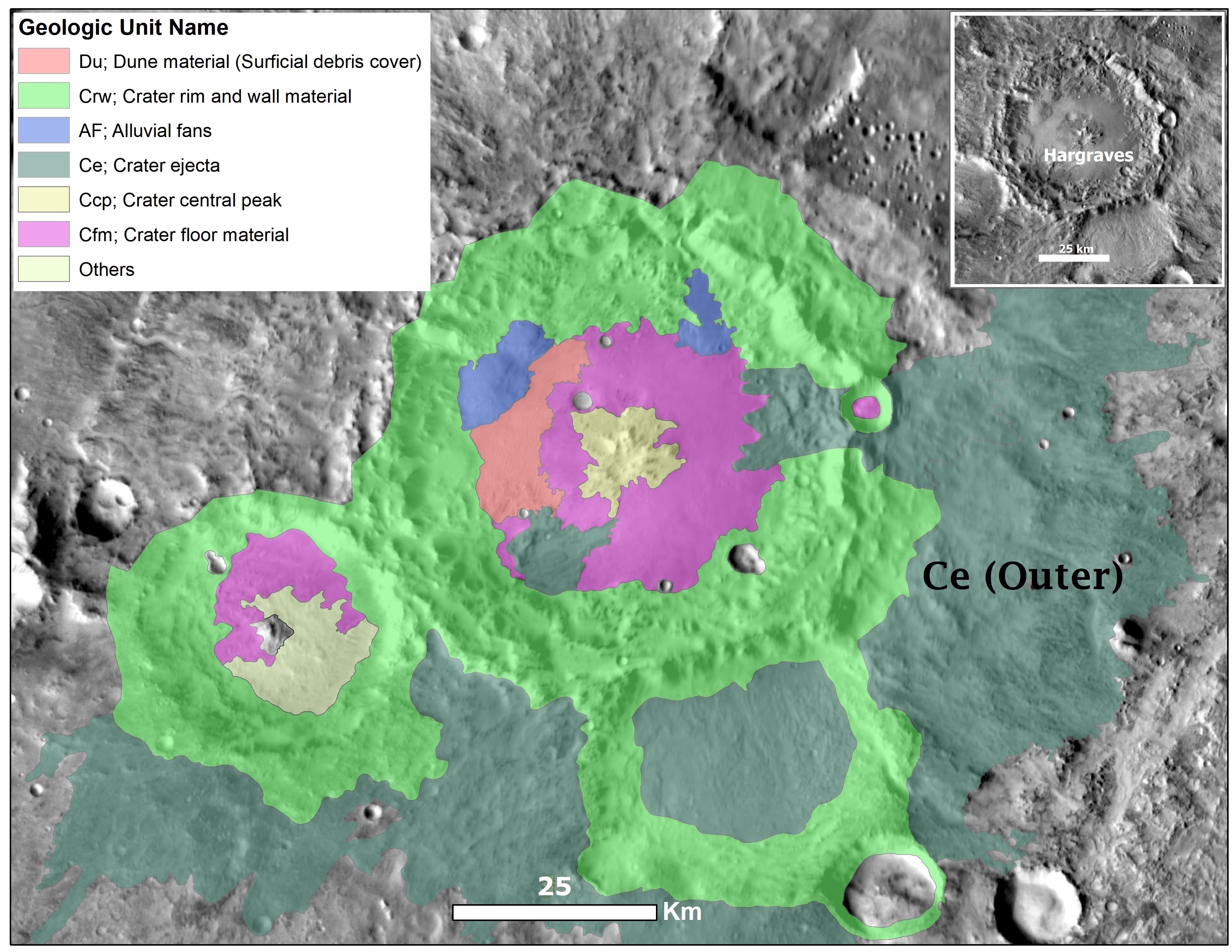}
\caption{A geologic map of Hargraves crater and its surroundings; crater ejecta (Ce), crater floor materials (Cfm), crater central peak (Ccp), rim and wall materials (Crw), alluvial fans (AF), and dune materials (Du) units within the crater and outside-crater ejecta (Ce (Outer)) unit. Inset (upper right) is the reference location. The background mosaic consists of THEMIS daytime infrared images. Note that the dune unit has been described as the surficial debris cover (Ac) unit in \cite{goudge2015assessing}. The figure is recreated after \cite{goudge2015assessing}. North is up.  \label{fig:02}}
\end{figure*}

Hargraves crater is a 65 km diameter impact crater centered at 75.75${}^{{}^\circ }$ E, 20.75${}^{{}^\circ }$ N, east of Nili Fossae (Fig. 1a), and formed during the late Noachian or early Hesperian \citep{ivanov2012major}.  The impact event that created Hargraves crater involved an array of different target bedrock lithologies, including a diverse set of mineralogy, such as mafic minerals and phyllosilicates \citep{mangold2007mineralogy}. Based on tone, texture, and morphological characteristics, \cite{goudge2015assessing} identified a few major morphologic units inside the crater such as an alluvial fan (AF), crater central peak (Ccp), crater floor materials (Cfm), crater ejecta (Ce), and surficial debris cover (Ac) (Fig. \ref{fig:02}). There is an outside-crater ejecta unit, hereafter we refer as Ce (Outer) unit. A dune field, visible at CTX and THEMIS resolutions, is located on the western side of the crater floor (Fig. \ref{fig:01}b). The dune field is labeled as surficial debris cover (Ac) and is characterized by aggregates of dunes \citep{goudge2015assessing}. The MGD${}^{3}$ and previous studies \citep{emran2019surficial, emran2019a, emran2020semiautomated} report that the dune field has barchan (B) and barchanoidal (Bd) dune types (see Fig. \ref{fig:03}b)\citep{mckee1979study}. Northwest of the dune field is an alluvial fan that was emplaced by fluvial activities along the crater rim \citep{mangold2007mineralogy}.
 
%%%%%%%%%%%%%%%%%%%%%%%%%%%%%%%%%%%%%%%%%%%%%%%%%%%%%%%%%%%%%%%%%%
\section{Observation and methods} \label{sec:method}
The bulk mineralogy was derived from TES data at spatial resolution roughly 3 x 6 km${}^{2}$. We use THEMIS ($\mathrm{\sim}$100 m/pixel) data for additional compositional analysis. Though TES has lower spatial resolution compared to THEMIS, it has many more spectral bands than THEMIS, which allows bulk mineral assemblages to be modeled. THEMIS images were used to examine grain size distribution using thermal inertia, as well as decorrelation stretch (DCS) as an additional compositional component.

\subsection{Thermal Emission Spectrometer (TES)} \label{subsec:tes}
TES, onboard the Mars Global Surveyor (MGS) spacecraft, has an infrared (5.8 to 50 $\mu$m) interferometric spectrometer, broadband thermal (5.1 $\mu$m to 150 $\mu$m) radiometer, and a visible/near-infrared (0.3 $\mu$m to 2.9-$\mu$m) radiometer \citep{christensen2001mars}. We use TES data recorded between 1651 cm${}^{-1}$ and 201 cm${}^{-1}$ with a sampling interval of 10 cm${}^{-1}$ resulting in 143 channels. To ensure the best quality data used in this study, we extract TES data following the method described in \cite{rogersbandfeild2009} and \cite{salvatore2018bulk}. TES spectra were extracted from early in the mapping orbit phase (orbits 1- 5317) that belong to high surface temperatures ($\mathrm{\ge}$260~K), low total ice ($\mathrm{\le}$0.04), and low total dust cover ($\mathrm{\le}$0.15). This quality control practice has successfully been applied in previous studies \citep{rogersChris2007, rogersetal2007, salvatore2014dominance, salvatore2016geologic} and helps to ensure the accuracy of the true spectral characteristics \citep{salvatore2018bulk}. 

Individual spectra were collected only from TES footprints that fall over the dune field, hereafter as dune unit (Du). For compositional comparison to inferred provenances, spectra were also collected over the crater floor materials (Cfm), crater ejecta (Ce), crater central peak (Ccp), and crater rim and wall materials (Crw) units. To ensure the spectra represented individual geologic units, data were only collected when TES footprints were fully within the bounds of a unit. We limit our spectra collection to between orbit counter keeper (OCK) 1583 and 7000 to account for instrumental error. Based on measured slipface direction, \cite{emran2020semiautomated} reported that the prevailing wind inside the crater flows to the west-northwest. Since the alluvial fan (AF) is located downwind of the dune field, we did not consider the alluvial fan unit (AF) as a potential source for the dunes. A list of TES spectra used for each geologic unit is given in appendix section of this paper (A1).

%%%%%%%%%%%%%%%%%%%%%%%%%%%%%%%%%%%%%%%%%%%%%%%%%%%%%%%%%%%%%%%%%%%%%%%%
\subsubsection{Spectral data processing}
TES spectral data were extracted using JMARS \citep{christensen2009jmars}. The data were then processed and analyzed with DaVinci programming environment \citep{edwards2015processing}. Prior to averaging the spectra for each representative morphologic unit, an atmospheric correction is applied to each orbital group (OCKs) spectra using the deconvolution algorithm of \cite{bandfield2000spectral}, \cite{smith2000separation}, and \cite{rogersahranson2008}. We used a library of atmospheric end-members \citep{bandfield2000spectral} and treated each spectra individually for atmospheric correction, which involves extracting surface emissivity through subtracting atmospheric contributions from measured emissivity spectra \citep{salvatore2018bulk}. Upon atmospheric removal, the surface emissivity spectra for each representative morphologic unit were averaged. The average surface composition and mineral assemblage were then modeled using an unmixing algorithm \citep{ramseychristian1998mineral, rogersahranson2008} with a selected mineral endmember library (see Table \ref{tab:endmem}). 

The algorithm uses a non-negative least square minimization routine \citep{rogersahranson2008} ~between the wavelength range of 230~cm${}^{-1}$~and 1305~cm${}^{-1}$. The non-negative least square routine employs all spectra in design matrix and remain available to the algorithm until a final non-negative solution is reached \citep{rogersahranson2008}. As the solution of the routine converges toward the best fit, a positive or zero spectral coefficient is maintained in the algorithm. However, the algorithm allows atmospheric and blackbody end-member coefficients to be negative. Details of the algorithm can be found in \citep{rogersahranson2008}. Using the spectral unmixing routine, we generated 8 mineral groups \citep{rogersfergonson2011} for each unit with a $\mathrm{\sim}$10\% accuracy and detection threshold for mineral abundance \citep{feelychristian1999, christensen2000thermal, bandfield2002global, rogersetal2007}.  The final outputs of spectral unmixing include modeled abundance mineral groups (feldspar, high-silica phases, pyroxene, olivine, hematite, sulfate, carbonate, and quartz), error estimation, and an individual end-member assemblage \citep{rogersfergonson2011}.

\startlongtable
\begin{deluxetable*}{lccc}
\tablenum{1}
\tablecaption{The list of thermal infrared spectral endmembers (both a library of atmospheric end-members and a library of surface end-member spectra) used for unmixing TES data in this study.\label{tab:endmem}}
\tablewidth{0pt}
\tablehead{
\colhead{Mineral Group} & \colhead{Spectrum/endmember name} &
\colhead{Mineral ID} & \colhead{Reference}
}
\decimalcolnumbers
\startdata
Quartz & Quartz  & BUR-4120 & [1] \\
Alkali feldspar & Microcline  & BUR-3460 & [1] \\ 
Plagioclase  & Albite  & WAR-0235 & [1] \\ 
 & Oligoclase  & BUR-060D & [1] \\
 & Andesine  & WAR-0024 & [1] \\
 & Labradorite  & BUR-3080A & [1] \\
 & Bytownite  & WAR-1384 & [1] \\
 & Anorthite  & BUR-340 & [1] \\
 & Shocked anorthosite & None & [2] \\
 &	at 17 GPa & & \\
 &	at 21 GPa & & \\
 &	at 25.5 GPa & & \\
 &	at 27 GPa & & \\
 &	at 38 GPa & & \\
 &	at 56.3 GPa & & \\
Orthopyroxene  & Bronzite  & NMNH-93527 & [1] \\ 
 & Enstatite  & HS-9.4B & [1] \\ 
 & Hypersthene  & NMNH-B18247 & [1] \\
Low-Ca clinopyroxene & Average Lindsley pigeonite & None & [3] \\
High-Ca clinopyroxene  & Diopside  & WAR-6474 & [1] \\ 
 & Augite  & NMNH-9780 & [1] \\
 & Augite  & NMHN-122302 & [1] \\
 & Hedenbergite manganoan  & DSM-HED01 & [1] \\
Olivine\newline  & Forsterite  & BUR-3720A & [1] \\
 & Fayalite  & WAR-RGFAY01 & [1] \\
 & Olivine Fo${}_{60}$ & KI 3362 & [4] \\
 & Olivine Fo${}_{68}$ & KI 3115 & [4] \\
 & Olivine Fo${}_{35}$ & KI 3373 & [4] \\
 & Olivine Fo${}_{10}$ & KI 3008 & [4] \\
Phyllosilicates  & Illite Imt-1 $\mathrm{<}$ 0.2~$\mu$m (pellet) & Imt-1 & [5] \\
 & Ca-montmorillonite solid & STx-1 & [1] \\
 & Saponite (Eb-1) $\mathrm{<}$ 0.2~$\mu$m (pellet) & None & [6] \\
 & SWy-1 $\mathrm{<}$ 0.2 microns (pellet) & None & [5] \\
Glass  & K-rich glass & None & [3] \\
 & SiO2~glass & None & [3] \\
Amorphous silica  & Opal-A  & 01--011 & [7] \\
 & Al-Opal & None & [8] \\
Oxide & Average Meridiani and Aram Hematite (TT derived) & None & [9] \\ 
Sulfate  & Anhydrite  & ML-S9 & [1] \\
 & Gypsum  & ML-S6 & [1] \\
Carbonate & Kieserite & None & [10] \\
 & Calcite  & C40 & [1] \\
Zeolite & Dolomite  & C20 & [1] \\
 & Crystalline heulandite & None & [11] \\
 & Crystalline stilbite & None & [11] \\
Atmosphere & Low-opacity dust & None & [12] \\
 & High-opacity dust & None & [12] \\
 & Water ice (small) & None & [12] \\
 & Water ice (large) & None & [12] \\
 & Synthetic CO${}_{2}$ & None & [12] \\
 & Synthetic water vapor & None & [12] \\
\enddata
\tablecomments{ [1] \cite{christensen2000thermal}; [2] \cite{johnson2002thermal}; [3] \cite{wyatt2001analysis}; [4] \cite{koeppenhamilton2008global}; [5] \cite{michalski2006emission}; [6] \cite{michalski2005mineralogical}; [7] \cite{michalski2003thermal}; [8] provided by M. D. Kraft/cited in \cite{rogersfergonson2011}; [9] \cite{glotch2004effect}; [10] \cite{baldridge2008thermal}; [11] \cite{ruff2004spectral}; [12] \cite{bandfield2000spectral}.}
\end{deluxetable*}

%%%%%%%%%%%%%%%%%%%%%%%%%%%%%%%%%%%%%%%%%%%%%%%%%%%%%%%%%%%%%%%%%%%%%%%%
\subsubsection{Regression models} \label{subsubsec:models}
We use linear regression to model the relationship between the bulk composition of each surrounding geologic unit and dune field by fitting a linear equation to the mineral abundance. Since spectral unmixing generates abundances of 8 mineral groups i.e., endmember abundance with a mean ($\mu$) and uncertainty (1-$\sigma$) value \citep{rogersfergonson2011}, we resample each endmember assuming the bulk composition as a gaussian scatter. We draw a total of 1,000 normal random samples (125 for each endmember) from each surrounding geologic unit centered around $\mu$ within the limit imposed by the 1-$\sigma$ of the endmember abundance. In our linear regression model, we assume the abundance of minerals in the dune field as the response (\textit{y}) and each surrounding geologic unit as an explanatory variable (\textit{x}). For gaussian scatter, the linear regression model uses the equation as:
\begin{equation} \label{eq1} 
y=ax+b 
\end{equation} 

where the slope of the line is~\textit{a,} and\textit{~b}~is the intercept or offset. Both \textit{a} and\textit{ b} are the parameters to estimate in the model. Using this model, we estimate the posterior probability distribution of the model parameters considering the uncertainty (1-$\sigma$ value) of the response (dune unit) using the Bayesian inference approach. The posterior probability distribution of the parameters of the linear model tells us the degree of similarity of the bulk mineralogy between the composition of each surrounding geologic unit to the dune field. Thus, the posterior probability distribution of the parameters indicates the likelihood of contribution of the bulk mineralogy from each geologic unit to the dune field’s bulk composition.

 We employ a multiple linear regression model to estimate how much of each surrounding unit mixing with other units could produce the dune field composition. For that, we assume the volumetric (bulk) percent contribution of mineral compositions from each of the surrounding units to the dune field. We adopt this approach as it is the simplest and physically possible scenario for the compositional mixing of geologic materials on Mars. In the case of volumetric percent contribution, the bulk compositions of the surrounding individual units (\textit{X${}_{i}$}) are summed with weights equal to the fractional contributed by each unit (\textit{F${}_{i}$}), as given below:

\begin{equation} \label{eq2} 
Y=\sum_{i}^{n}{F_iX_i} + \textit{Z${}_{0}$}
\end{equation} 

where 0  $\leq$ \textit{F${}_{i}$} $\leq$ 1, \textit{Y} is the bulk composition of the dune field, \textit{Z${}_{0}$} is the additive offset, and \textit{i} represents the surrounding individual unit. In our model, the fractions contributed by each unit (\textit{F}${}_{i}$) and \textit{Z${}_{0}$ }are the parameters to estimate considering the uncertainty (1-$\sigma$) of \textit{Y}. We estimate the posterior probability distribution of these parameters of the multiple linear model which tells us the fractional contribution of bulk mineralogy from each surrounding unit to the dune field composition. The model uses a non-negative linear combination (mixing) strategy i.e., \textit{F}${}_{i}$ must render a positive value.

We use a Markov Chain Monte Carlo (MCMC) simulation technique \citep{hogg2018data} to estimate the parameters of our models. Inherently the MCMC employs a Bayesian inference approach for sampling the probability density function, performing probabilistic inferences, and fitting models to data \citep{hogg2018data}. For detail of the mathematical foundation of the MCMC technique refer to \cite{goodmanweare2010}; \cite{hogg2018data}. We use {\tt\string emcee} \citep{foreman2013emcee}, a Python routine for MCMC simulation proposed by \cite{goodmanweare2010}, for the implementation of our models. With the MCMC routine, we use 1,000 iterations to estimate the model parameters. We report the median, including 16\% quantile as lower 1-$\sigma$ error and 84\% quantile as an upper 1-$\sigma$ error, fraction contribution of the bulk composition of each surrounding unit using a corner plot diagram.

%%%%%%%%%%%%%%%%%%%%%%%%%%%%%%%%%%%%%%%%%%%%%%%%%%%%%%%%%%%%%%%%%%%%%%%%
\subsection{Thermal Emission Imaging System (THEMIS)} \label{subsec:THEMIS}
THEMIS consists of two multispectral imagers onboard the 2001 Mars Odyssey spacecraft, enabling a useful analysis of the thermophysical, composition, and physical properties of the martian surface \citep{christensen2004thermal}. The THEMIS sensor consists of visible (VIS), near-infrared (NIR), and thermal-infrared (TIR) imagers. The thermal infrared sensor has 10 channels ranging from 6.78 $\mu$m to 14.88 $\mu$m at a spatial resolution for daytime and nighttime images at 100 m/pixel \citep{christensen2004thermal}. We used THEMIS daytime IR to analyze surface composition and nighttime IR to analyze thermophysical characteristics.

Surface compositional and thermal inertia data can be extracted from representative surface units through overlapping THEMIS daytime and nighttime IR images. This allows an assessment of the relationship between composition and grain-size distribution \citep{bandfield2011role}. THEMIS daytime radiance data are atmospherically corrected, calibrated, and converted to surface emissivity using the methods described in \cite{christensen2004thermal}, \cite{bandfield2004a}, and \cite{edwards2011mosaicking}. We used the THEMIS I02781003 image acquired under an average surface temperature of 247${}^{\ }$K, water ice opacity of 0.072, and low atmospheric dust opacity of 0.052 at 9 $\mu$m. THEMIS images with a similar specification of average surface temperature and low atmospheric dust and water ice opacity have successfully been used in previous studies \citep{bandfield2004b}.

%%%%%%%%%%%%%%%%%%%%%%%%%%%%%%%%%%%%%%%%%%%%%%%%%%%%%%%%%%%%%%%%%%%%%%%%
\subsubsection{Thermal inertia}
Thermal inertia (TI) is considered as a primary physical property measuring the resistance of materials to a temperature change and can be used as an indicator of surface geological characteristics for the upper few centimeters \citep{jakosky2000thermal, mellon2000high, putzig2005global}. Thermal inertia is influenced by thermal conductivity and surface physical structure such as particle grain-size, induration, porosity, rock abundance, bedrock exposure, and compaction of surface materials \citep{kieffer1973preliminary, putzig2005global, fergason2006high, piqueuxchristian2011, putzig2014}. Thermal inertia of material, \textit{I }(J m${}^{-2}$ K${}^{-1}$s${}^{-1/2}$) can be defined as:
\begin{equation} \label{eq3} 
I=\sqrt{}(k\rho c) 
\end{equation} 

where \textit{k} is the thermal conductivity (W m${}^{-1}$ K ${}^{-1}$), \textit{$\rho$} is the density (kg m${}^{-3}$), and \textit{c} is the specific heat (J kg${}^{-1}$ K${}^{-1}$). On Mars, the value of \textit{$\rho$c} is assumed to be 1 x 10${}^{6}$ Jm${}^{-3}$K${}^{-1}$ \citep{neugebauer1971mariner, fergason2006high}. 

Ideally, higher thermal inertia values are associated with compacted rock and/or larger particle-size materials corresponding to a mechanically strong surface, such as well-indurated (cemented) rock. In contrast, weakly lithified rocks and/or unconsolidated (weakly indurated) materials are associated with lower thermal inertia values \citep{williams2018}. Thus, lower thermal inertia ($\mathrm{<}$ $\mathrm{\sim}$350 Jm${}^{-2}$K${}^{-1}$s${}^{-0.5}$) represents loose and fine-grained surface particles \citep{piqueuxchristian2009a, piqueuxchristian2009b} whereas higher thermal inertia ($\mathrm{>}$ 1,200 Jm${}^{-2}$K${}^{-1}$s${}^{-0.5}$) indicates igneous bedrock \citep{edwards2009global}. We extracted quantitative thermal inertia from the THEMIS global thermal inertia product \citep{christensen2013themis} produced using the method of \cite{fergason2006high}. The overall accuracy and precision of THEMIS-derived thermal inertia is around $\mathrm{\sim}$ 20\% \citep{fergason2006high}. 

Effective grain-size can be interpreted from thermal inertia values \citep{kieffer1973preliminary}. We calculated the effective particle size diameter of the dune materials using the following equation:
\begin{equation} \label{eq4} 
k=\left(C*P^{0.6}\right)*d^{-0.11.log\ (\frac{P}{K})} 
\end{equation} 

where \textit{C} and \textit{K} are constants, with values of 0.0015 and 8.1 x 10${}^{4}$ torr, respectively, \textit{P} is the atmospheric pressure in torr, \textit{d} is the average particle diameter measured in $\mu$m, and \textit{k} is the thermal conductivity in Wm${}^{-1}$K${}^{-1}$, which was derived using equation \ref{eq3} \citep{presleychristian1997b, presleychristian2010}. Mars' average surface pressure of 3.9 torr (5.2 mbar or 520 Pa) at L${}_{s\ }$= 0$\mathrm{{}^\circ}$ \citep{smithzuber1998} was used as the reference atmospheric pressure in this study. Using equation (\ref{eq4}), the derivation of particle diameter is valid for thermal inertia values less than $\mathrm{\sim}$350 and an expected precision of the measurements in particle size is to be $\mathrm{\pm}$10-15\% \citep{presleychristian1997b, fergason2006high}. Thermal inertia values higher than 350 are difficult to interpret and may be indicative of a solid rock outcrop mixed with fine-grained material within the footprint of thermal data \citep{presleychristian1997b}.

%%%%%%%%%%%%%%%%%%%%%%%%%%%%%%%%%%%%%%%%%%%%%%%%%%%%%%%%%%%%%%%%%%%%%%%%
\subsubsection{Decorrelation stretch (DCS)} \label{subsubsec:dcs}

Decorrelation is a contrast stretch method that displays enhanced information from multispectral thermal infrared images. Refer to \cite{gillespie1986color} for details. We use DCS images \citep{edwards2011mosaicking} derived as a standard THEMIS IR radiance product \citep{bandfield2004b} to highlight surface spectral and compositional variability at a spatial resolution much higher than TES. DCS images are typically rendered with 3-band combinations such as 8-7-5, 9-6-4, or 6-4-2 displayed in red, green, and blue channels, respectively. Yellow appearance in DCS images using bands 8-7-5 and 9-6-4, and magenta using bands 6-4-2 tend to correspond with an elevated bulk-silica content \citep{amador2016elevated}. Materials that appear purple in both 8-7-5 and 9-6-4 DCS images and cyan in 6-4-2 typically represent olivine-bearing basalts \citep{edwards2011mosaicking, bandfield2011role, amador2016elevated, salvatore2016geologic}. We primarily focus on DCS images stretch using bands 8-7-5- in red, green, and blue channels, respectively, to highlight compositional variability present within the dune field and surrounding areas, because this band combination exhibits most variability. The color variability in DCS images is scene-specific and the spectral variability of DCS within a particular scene should be interpreted for that specific scene only because of its temperature dependence  \citep{rogersfergonson2011}.

%%%%%%%%%%%%%%%%%%%%%%%%%%%%%%%%%%%%%%%%%%%%%%%%%%%%%%%%%%%%%%%%%%%
\section{Results} \label{sec:results}
The result of the grain-size distribution for the dune materials was presented first followed by the results of dune morphology and stability, and bulk mineralogy of the dune field and the potential provenances. Then the result of bulk mineralogy from the TES analysis was revisited using THEMIS-derived analysis of DCS image.

\subsection{Grain-size distribution}
\begin{figure*}[ht!]
\plotone{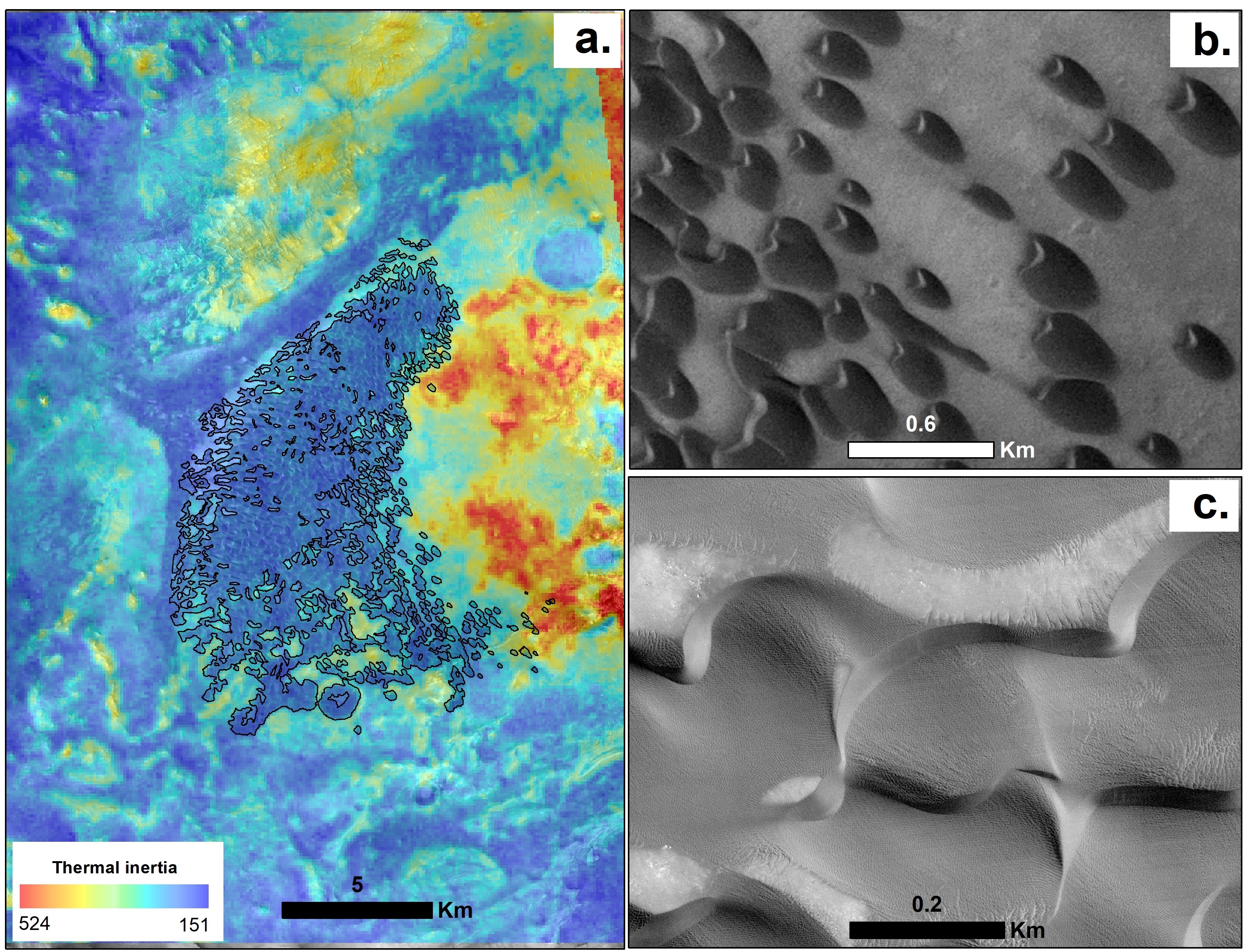}
\caption{a) Thermophysical characteristics of the dunes and surrounding materials derived from nighttime THEMIS thermal infrared data overlain onto a CTX mosaic. Warmer colors represent higher thermal inertia. Over the dunes and the associated sand aprons, there is comparatively lower thermal inertia than over the surrounding areas. b) Tonal variation of dunes from non-dune materials in CTX image resolution where darker tone distinguishes dunes from non-dune materials. c). Distinctly identifiable crests and slipface are visible in high-resolution HiRISE image. North is up.  \label{fig:03}}
\end{figure*}

THEMIS nighttime data are more reliable than daytime data when deriving thermal inertia and estimating grain-size distribution because nighttime observation minimizes the effects of slope and albedo \citep{edwards2018thermophysical}. Quantitative thermal inertias were extracted using the THEMIS global thermal inertia mosaic by \cite{christensen2013themis} for the dune field. The estimated average thermal inertia across the dune field is 238$\mathrm{\pm}$17 \footnote{Mean $\mathrm{\pm }$ 1-$\mathrm{\sigma}$ standard deviation} Jm${}^{-2}$K${}^{-1}$s${}^{-0.5}$, with a minimum value of 200 Jm${}^{-2}$K${}^{-1}$s${}^{-0.5}$ and a maximum of 426 Jm${}^{-2}$K${}^{-1}$s${}^{-0.5}$. This value is consistent with the average thermal inertia measured in the Bagnold dune field using multiple THEMIS observations \citep{edwards2018thermophysical}. The thermal inertia map (Fig. \ref{fig:03}a) shows a concentration of relatively similar values within the outline of the dune field. Visual inspection indicates a comparatively lower thermal inertia across the dune field when compared to the surrounding geologic units. For instance, the crater floor materials (Cfm) on the east side of dunes have an average thermal inertia value of $\mathrm{\sim}$400 Jm${}^{-2}$K${}^{-1}$s${}^{-0.5}$.

We calculate the average particle size across the dune field using the equations (\ref{eq3} - \ref{eq4}). We convert each THEMIS thermal inertia pixel to particle size instead of converting average thermal inertia to average particle size. We adopted this method because the conversion of thermal inertia to particle-size is non-linear \citep{edwards2018thermophysical}. The average grain-size values were calculated after converting each TI pixel to grain-size, which resulted in an average particle size of $\mathrm{\sim}$391$\mathrm{\pm}$172 $\mu$m. This particle size diameter corresponds to mostly medium sand-sized materials mixed with fine and coarse grain sands \citep{presleychristian1997a}. 

\subsection{Morphology and stability} \label{sec:morpho}
\begin{figure*}[ht!]
\plotone{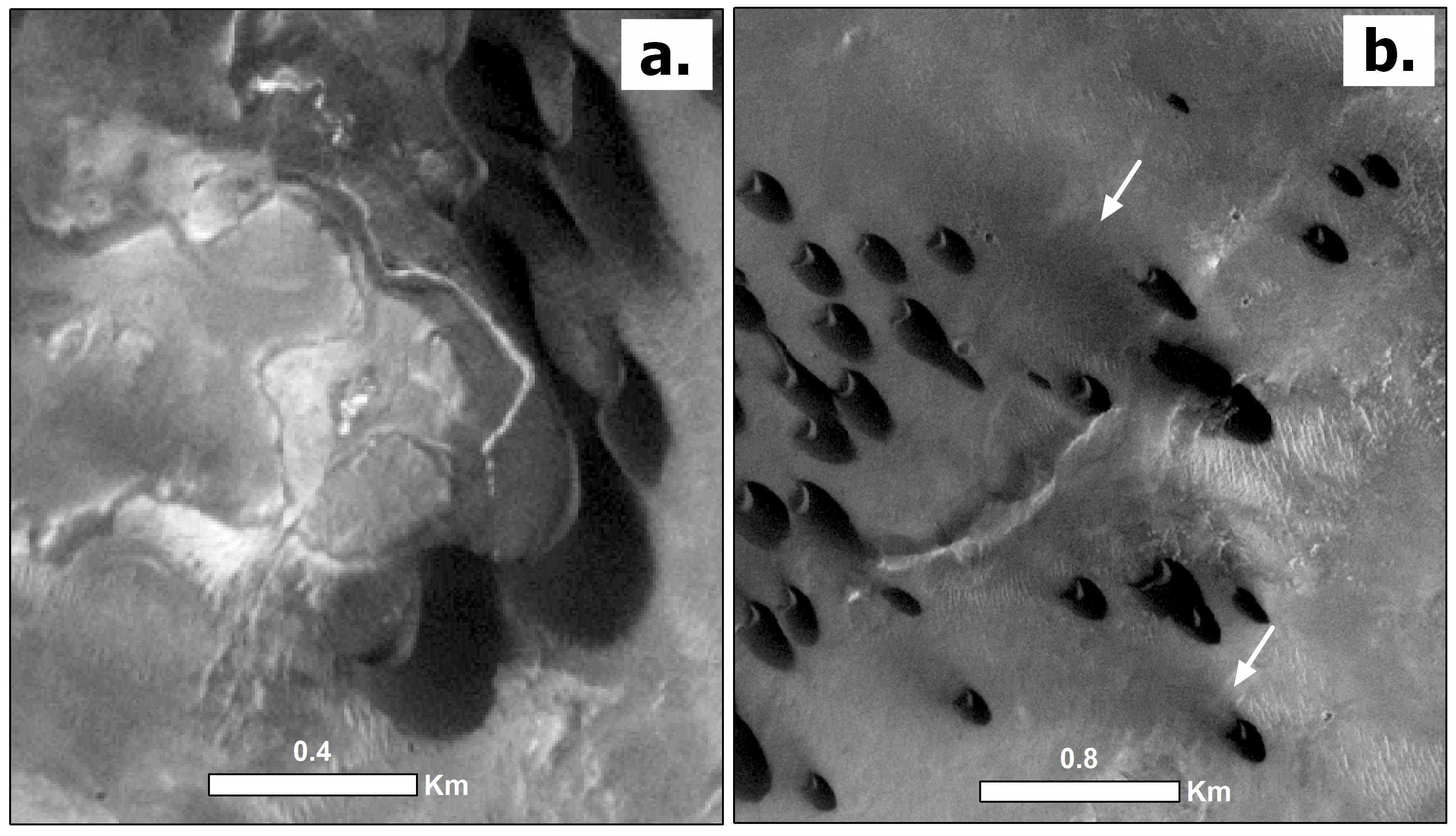}
\caption{a) An exposed light-toned feature in the south-eastern fringe from CTX image. The feature, interpreted to be a ``possible'' intrusion of igneous origin, has a semi-conical shape, appears to be layered, and dips away from the central point. b) Thin dark streaks, indicated by white arrows, emanating from the dunes on the south-eastern boundary of the dune field. The streaks are interpreted as the result of wind activities. North is up. \label{fig:04}}
\end{figure*}

Like the typical dunes elsewhere on Mars \citep{hayward2007amars, hayward2010mars, hayward2012mars, fenton2019bmars}, the dune field on the floor of Hargraves crater is darker in tone when compared to surrounding materials (Fig. \ref{fig:03}b-c). A consistent tone across the dune field suggests that dune surface materials are relatively homogenous. Visual inspection of high-resolution morphology from CTX and HiRISE images finds an absence of \textit{in-situ} bedrock surface within the dune field, except exposure of a light-toned feature in the south-eastern fringe (Fig. \ref{fig:04}a). The feature has a semi-conical shape and appears to be layered. The layers dip away from the central point and have been eroded to the extent that they show the layered nature of this feature. A central vent, which is in line with some well-defined fractures, shows lighter shaded materials residing below the eroded layers that enshroud the feature. The feature can be interpreted to be a ``possible'' intrusion of igneous origin, developed from small (isolated) volcanos \citep{richardson2021small}, supported by the nearby presence of the Syrtis Major volcanic province \citep{hiesingerhead2004syrtis}. However, this interpretation is based on our visual inspection and other possible interpretations of this feature may also be plausible. The physical relationship of this feature with surrounding dunes does not, unambiguously, reveal the relative age of the feature.

\begin{figure*}[ht!]
\plotone{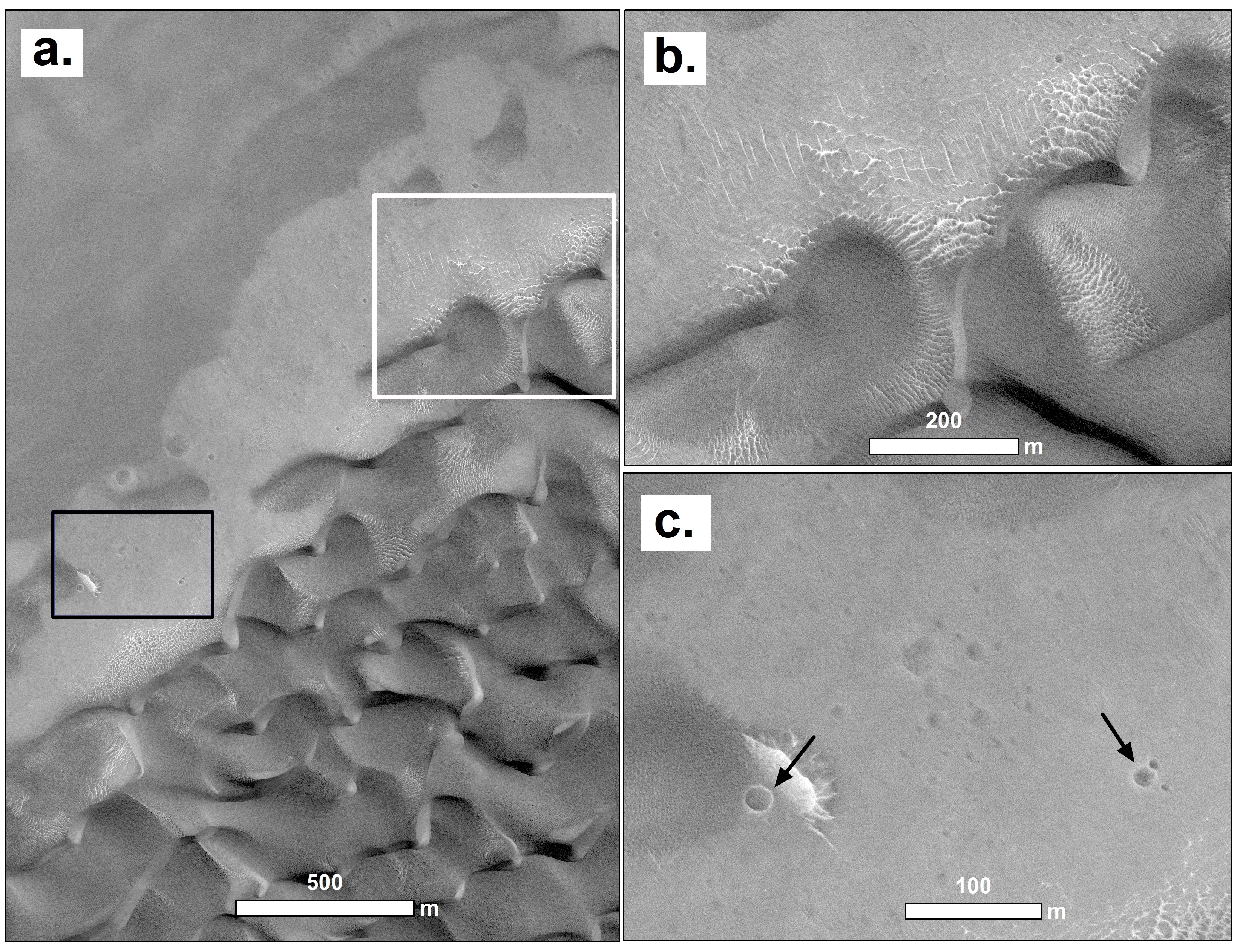}
\caption{a) Fine-scale morphology in the dune field from HiRISE ESP\_030302\_2010 image. The white and black rectangles (polygons) indicate the reference locations of b) and c), respectively. b) Transverse Aeolian Ridges (TARs) along the dune faces, near the crests. TARs are light-toned eolian mega-ripple features that appear to follow trellis and dendritic patterns. c) Small impact craters (black arrows) on mantling surface. North is up. \label{fig:05}}
\end{figure*}

We investigate fine-scale induration morphologies, non-eolian modification (or erosion), and degradation characteristics within the dune field, such as erosional features or impact cratering from HiRISE images ESP\_030302\_2010 and ESP\_030091\_2010 (Fig. \ref{fig:05}). An investigation of these images reveals a sand apron on the western boundary of the dune field. We also find potential mantling on the boundary between the dunes and sand apron, as well as impact cratering on the mantling material. The presence of potential mantling along the boundary which appears to expand into the dune field by filling in space between the dunes, small impact craters, and an average thermal inertia value of $\mathrm{\sim}$250 Jm${}^{-2}$K${}^{-1}$s${}^{-0.5}$ in the dune field, may be an indication of weak induration or lithification of the dune materials \citep{fergason2012surface}. We did not find other non-eolian or erosional features like gullies or mass wasting features. However, the dunes on the south-eastern boundary of the dune field show a different scenario. CTX (see Fig. \ref{fig:01}b and \ref{fig:04}b) image reveals thin dark streaks emanating from the dunes on the south-eastern boundary of the dune field.

There are features atop several dune crests and upper windward sides, such as sets of straight or very nearly straight grooves, akin to eolian ripples, of several different widths, orientations, and spacings. All the various sets of grooves have in common that they are roughly perpendicular to the crest of the dunes upon which they reside. There are also features upon the dunes including curved and scalloped grooves that follow trellis and dendritic patterns (Fig. \ref{fig:05}b). These curved and scalloped features have in common that they are closer to the crests of the dunes upon which they reside than are the nearby straight or nearly straight grooves, and in some instances persist down the leeward (avalanche) face of the dune. Both of these groove features exhibit shallow penetration into the dune structure and reveal lighter color sands. These bright-toned grooves are interpreted to be martian eolian mega-ripples, often refer as Transverse Aeolian Ridges-TARs \citep{silvestro2020megaripple}.

We assign the dune field within Hargraves carter a stability index of 2, since it has a partial apron that exists along the western side of the dune field \citep{fentonhayward2010southern}. Note that the dune field may overlap the morphological characteristics between multiple stability index classes and assignment to an appropriate class is a subjective procedure. Southern high-latitudes dunes show increased stability and inactivity \citep{fenton2019amars} due to presence of subsurface (near-surface ground) water ice that could stabilize dunes, and their thermal model generally fits to rock/ice thermal signature \citep{gullikson2018mars}. Unlike the southern-high latitude dune fields \citep{fentonhayward2010southern, gullikson2018mars, fenton2019amars}, the dunes within Hargraves crater may not be affected by subsurface ice or volatiles \citep{emran2020semiautomated}. We presume this because the dunes at the crater are potentially active and deprived of stability features and morphologies- indicating the absence of influences from subsurface (near-surface ground) water ice that would otherwise stabilize the dunes. Subsurface ice extends as far equatorward as 45$\mathrm{{}^\circ}$S between 40 and 140$\mathrm{{}^\circ}$ E longitudes \citep{fentonhayward2010southern}. Martian water-equivalent hydrogen-rich deposits of $\mathrm{\sim}$ $\mathrm{>}$20\% (by mass) at poleward of $\mathrm{\pm}$50$\mathrm{{}^\circ}$ latitudes and less rich deposits near-equatorial latitudes were reported from Neutron data observed using the Neutron Spectrometer aboard 2001 Mars Odyssey \citep{feldman2004global}. Thus, residing in the equatorial region ($\mathrm{\sim}$20${}^{{}^\circ }$ N), he dunes at Hargraves are less likely to be affected by subsurface ice or volatiles.

\subsection{Bulk composition}

We model bulk mineralogy across the dune field (Du) and surrounding geologic units within the crater. We then compare the dune composition with potential provenances: crater floor materials (Cfm), crater ejecta (Ce), central peak (Ccp), and crater rim and wall (Crw) units. The bulk composition analyzed here consists of different mineral groups \citep{rogersfergonson2011} based on the spectral library used in the unmixing analysis from TES emissivity data. Spectral unmixing generates eight mineral groups, such as feldspar (both plagioclase and alkali feldspar), pyroxene (both high-Ca clinopyroxene [HCP] and low-Ca clinopyroxene [LCP]), high-silica phase (HSP; including amorphous silica and phyllosilicates), sulfate, olivine, hematite, carbonate, and quartz. Our compositional modeling results are presented in Table \ref{tab:result}. Individual mineral endmember abundances are listed in appendix section of this paper (A2). We report the unmixing result of model abundances to nearest integer value and model error to hundredths decimal place as similarly reported by previous studies \citep{rogersChris2007}. Though the model produces mineral abundances out to several decimal places, TES unmixing results are prone to $\mathrm{\sim}$10\% uncertainly \citep{rogersetal2007} regardless of the actual model error reported. Furthermore, the model error in TES unmixing results is strongly tied to mineral endmember selection. The precision of mineral abundances and model error we report in this study are based on what the unmixing model has produced.

\begin{deluxetable*}{lccccc}
\tablenum{2}
\tablecaption{Our spectral unmixing result for dune (Du), crater floor materials (Cfm), central peak (Cp), and crater rim and wall (Crw) unit within the crater as defined by \cite{goudge2015assessing}. Reported here are the average areal abundances of each mineral group (\%) along with the calculated model error (RMSE). All values were normalized to the blackbody (BB) percentage (\%).\label{tab:result}}
\tablewidth{0pt}
\tablehead{
\colhead{Mineral Group} & \colhead{Dune(Du)} & \colhead{Floor Materials (Cfm)} & \colhead{Crater ejecta (Ce)} & \colhead{Central peak (Ccp)} & \colhead{Rim and wall (Crw)}
}
\decimalcolnumbers
\startdata 
Feldspar & 36$\mathrm{\pm}$7 & 24$\mathrm{\pm}$6 & 39$\mathrm{\pm}$4 & 27$\mathrm{\pm}$6 & 36$\mathrm{\pm}$7 \\ 
Pyroxene & 22$\mathrm{\pm}$6 & 29$\mathrm{\pm}$4 & 29$\mathrm{\pm}$4 & 21$\mathrm{\pm}$5 & 21$\mathrm{\pm}$5 \\ 
Olivine & 12$\mathrm{\pm}$5 & 7$\mathrm{\pm}$5 & 3$\mathrm{\pm}$2 & 8$\mathrm{\pm}$4 & 7$\mathrm{\pm}$3 \\
High-silica phase & 16$\mathrm{\pm}$7 & 23$\mathrm{\pm}$5 & 15$\mathrm{\pm}$4 & 22$\mathrm{\pm}$6 & 21$\mathrm{\pm}$5 \\
Carbonate & 5$\mathrm{\pm}$1 & 3$\mathrm{\pm}$1 & 4$\mathrm{\pm}$1 & 4$\mathrm{\pm}$1 & 3$\mathrm{\pm}$1 \\
Sulfate & 3$\mathrm{\pm}$2 & 10$\mathrm{\pm}$1 & 7$\mathrm{\pm}$1 & 12$\mathrm{\pm}$3 & 9$\mathrm{\pm}$2 \\
Hematite & 4$\mathrm{\pm}$1 & 4$\mathrm{\pm}$1 & 4$\mathrm{\pm}$1 & 6$\mathrm{\pm}$2 & 6$\mathrm{\pm}$1 \\
Quartz & 1$\mathrm{\pm}$2 & 0$\mathrm{\pm}$0 & 0$\mathrm{\pm}$0 & 0$\mathrm{\pm}$0 & 1$\mathrm{\pm}$1 \\
Blackbody & 0.4$\mathrm{\pm}$3 & 0.4$\mathrm{\pm}$2 & 0.2$\mathrm{\pm}$2 & 0.4$\mathrm{\pm}$3 & 0.5$\mathrm{\pm}$2 \\
RMSE & 0.25 & 0.22 & 0.28 & 0.26 & 0.17 \\
\enddata
\end{deluxetable*}

\begin{figure*}[ht!]
\plotone{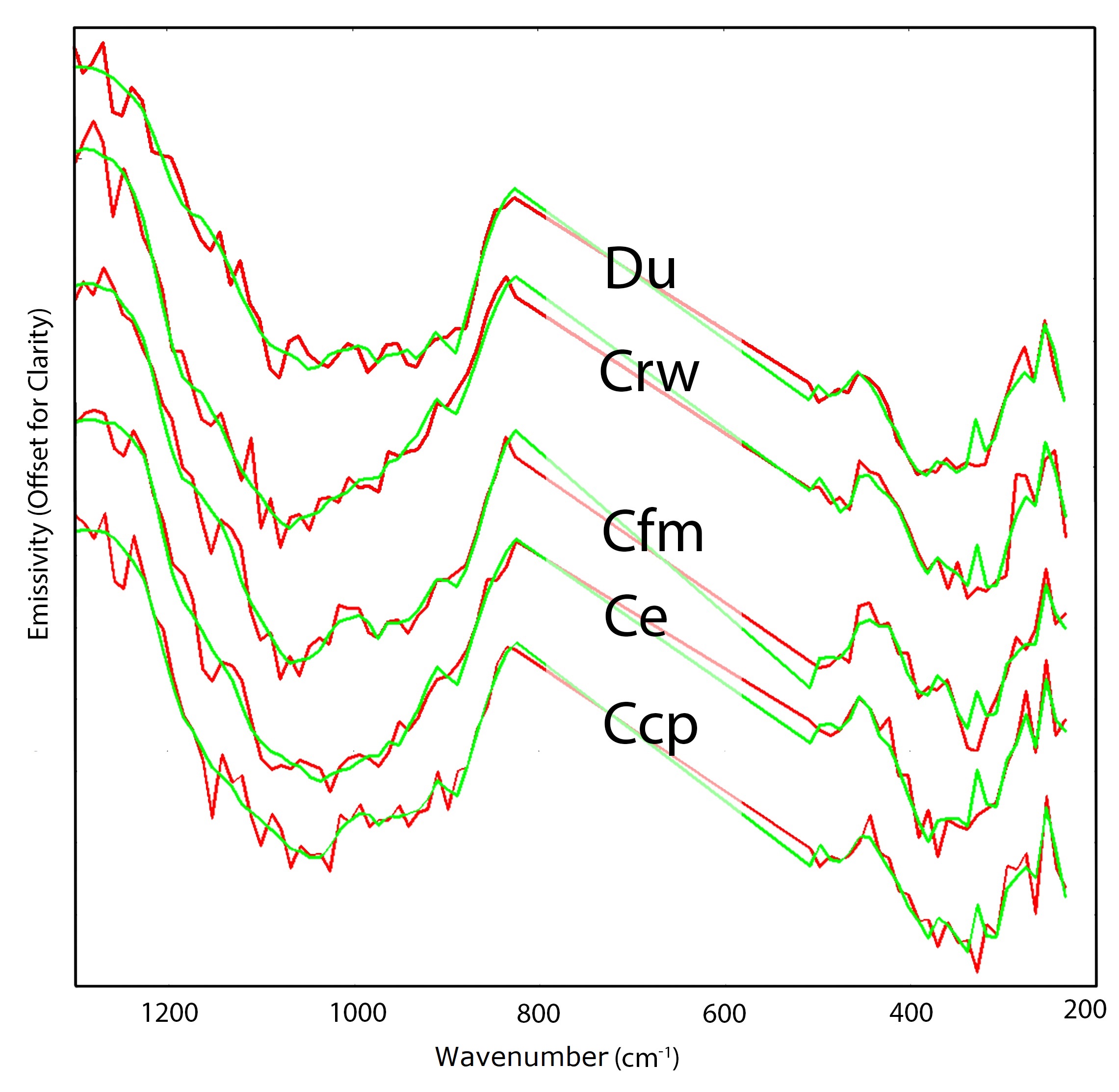}
\caption{Average atmospherically corrected TES~emissivity~spectra (red) for each of the investigated morphologic units in Hargraves crater, with the best-fit model spectra (green) derived from the linear unmixing procedures. For interpretation of the references to color in this figure legend, the reader is referred to the web version of this article. \label{fig:06}}
\end{figure*}

Our unmixing results reveal that the dune field is mainly comprised of a mixture of feldspar, pyroxene, olivine, and other silicate minerals. The Du unit has both the highest concentration of olivine (12$\mathrm{\pm}$5\%) and lower concentration of high-silica phase minerals (16$\mathrm{\pm}$7\%) compared to surrounding units. For example, the Ccp unit has the second-highest olivine abundance, modeled at (8$\mathrm{\pm}$4\%), but higher concentration of high-silica phase minerals (22$\mathrm{\pm}$6\%). The Ce unit has the highest concentration of feldspar (39$\mathrm{\pm}$4\%) but has the lowest abundance of olivine (3$\mathrm{\pm}$2\%). The Cfm unit has the highest abundances of high-silica phases (23$\mathrm{\pm}$5\%), but the lowest feldspar concentration (24$\mathrm{\pm}$6\%). The average surface emissivity spectra and modeled spectra for the five geomorphic units within the crater are shown in Fig. \ref{fig:06}.

\begin{figure*}[ht!]
\plotone{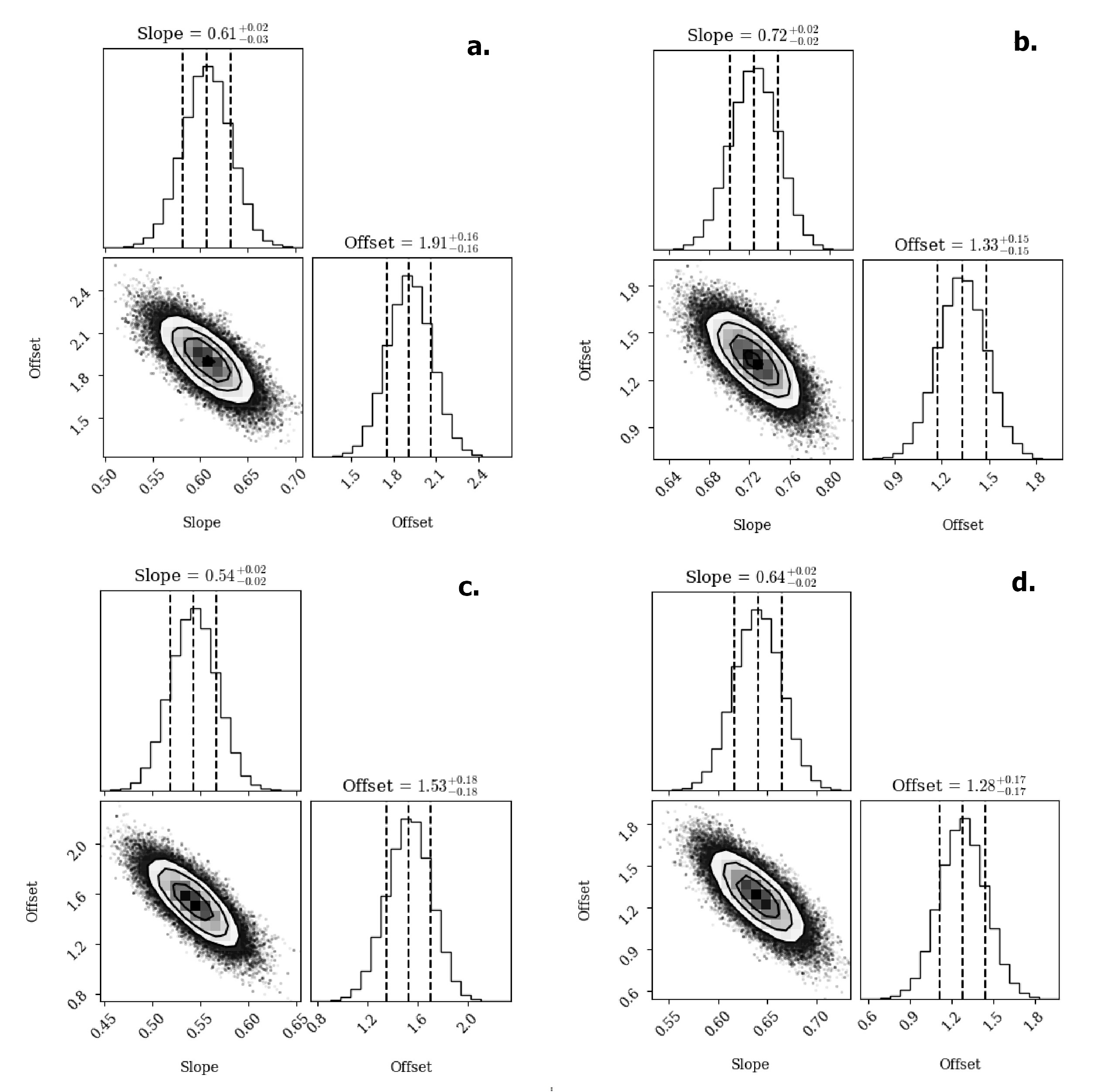}
\caption{Posterior probability distribution of the linear model parameters slope and offset for each unit within the crater such as a) Cfm, b) Ce, c) Ccp, and d) Crw. The number value represents the median, including their 16\% quantile as lower 1-$\sigma$ error bar and 84\% quantile as an upper 1-$\sigma$ error bar, of linear model's slope and offset for the respective geologic units. \label{fig:07}}
\end{figure*}

We model the relationship between the bulk mineralogic composition of the geological units within the crater and the dune field. Our results from the linear regression show the posterior probability distribution of slope and offset (intercept) of the model fit for each geologic unit (Fig. \ref{fig:07}). The values for the posterior probability distribution of the slope parameter in Fig. \ref{fig:07} indicate the degree of similarities, and thus the likelihood of contribution by, each of the surrounding units has to the dune field. The Ce unit has the highest median slope value of ${0.72}^{+0.01}_{-0.01}$, followed by the Crw (${0.64}^{+0.02}_{-0.02}$), Cfm ${(0.60}^{+0.01}_{-0.01}$), and Ccp ${(0.54}^{+0.01}_{-0.01}$) units. Based on the median slope values from the posterior probability distribution, we suggest that the likelihood of the bulk mineralogy in the Ce unit to the bulk composition of the dune field is the highest whereas the Ccp has the lowest likelihood within the crater.

\begin{figure*}[ht!]
\plotone{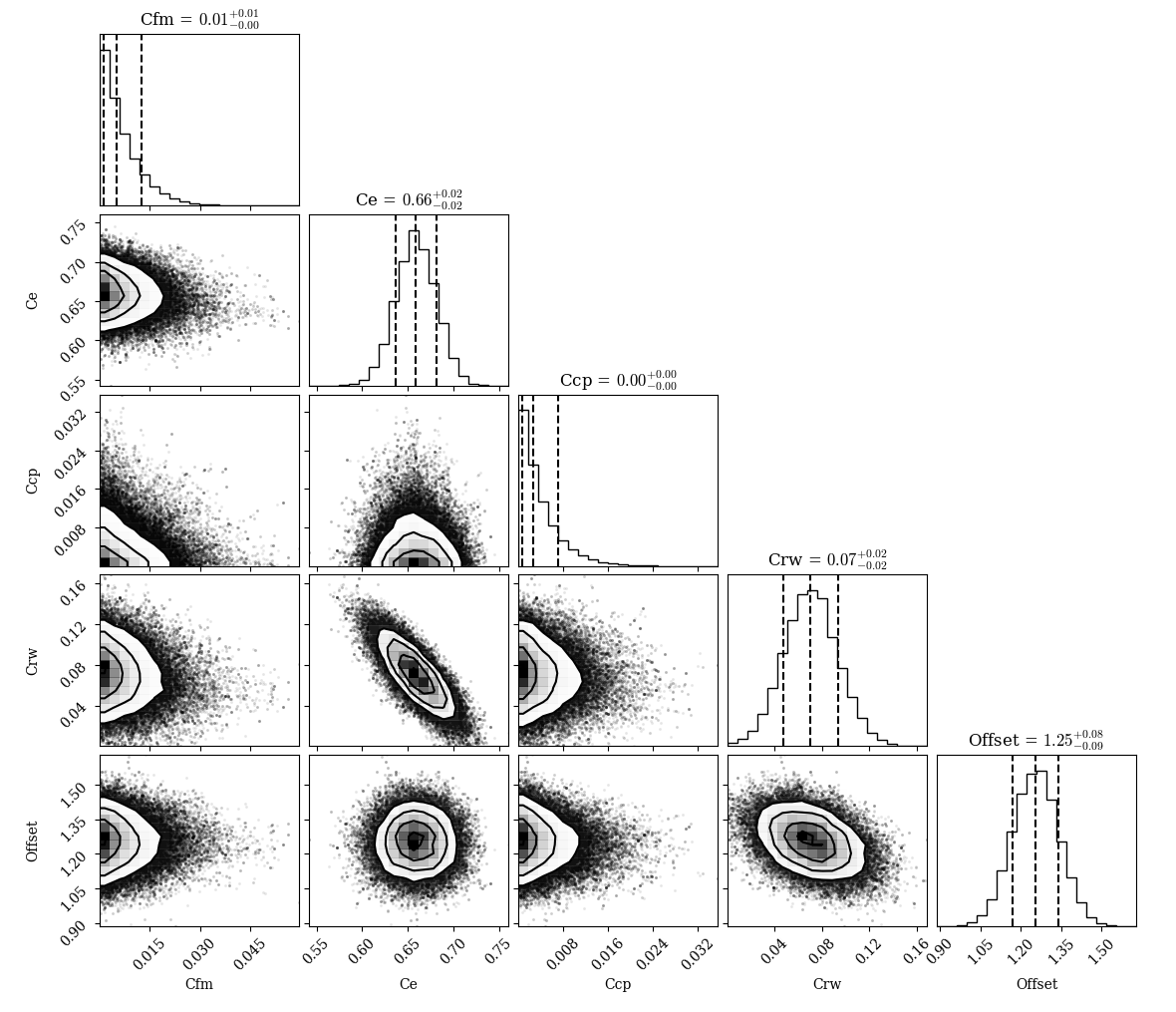}
\caption{Posterior probability distribution of the model parameters as the fractions contributed by each unit within the crater. The number value represents the median, including their 16\% quantile as lower 1-$\sigma$ error bar and 84\% quantile as an upper 1-$\sigma$ error bar, of fractions contributed by respective individual unit.\label{fig:08}}
\end{figure*}

We model the bulk compositions of the geological units within the crater to estimate the possible fraction amount contributed by each unit to the dune field. We show the result of the posterior probability distribution of the parameters of our multiple linear regression model in Fig. \ref{fig:08}. That is the values for the posterior probability distribution for each of the unit indicates the fractional contribution from each of the surrounding units to the dune field. The result of our modeled parameters show that the posterior probability distribution of the Ce unit has dominantly the highest median fraction of ${0.66}^{+0.02}_{-0.02}$, followed by a disproportionally lower fraction contributed by the Crw (${0.07}^{+0.03}_{-0.03})$ and Cfm (${0.01}^{+0.01}_{-0.00})$ units. Our model did not predict any fraction contribution by the Ccp unit as reported in Fig. \ref{fig:08}. This indicates that the dune materials are, likely, predominantly sourced from the adjacent Ce unit within the crater- consistent with the linear model result described above. The Crw and Cfm units within the crater contribute to the dune composition mixture at lower fractions. 

To assess the possibility of potential contribution from the geological units outside the crater, we further run MCMC simulation by including the outside-crater ejecta, Ce (Outer), unit located on the eastern and south-eastern sides of the crater (see Fig. \ref{fig:02}). We consider the Ce (Outer) unit because it is the largest geological unit, extents to around $\mathrm{\sim}$30 km from the crater rim and wall (Crw), outside the crater. The selection of this unit is also based on the dune field's raw slipface direction, indicating wind movement, and its upwind position in respect to the dune field. To minimize possible spectra selection-induced bias, we used a maximum number (24 observations) of TES spectra from the Ce (Outer) unit. We adopt the same spectral unmixing procedure to estimate the bulk composition of the Ce (Outer) unit.

\begin{figure}[h!]
\plotone{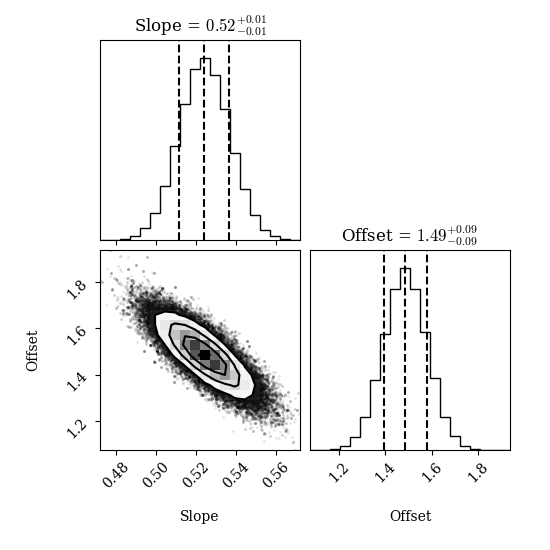}
\caption{Posterior probability distribution of the linear model parameters as slope and offset for the Ce (outer) unit and dune field. The number value represents the median, including their 16\% quantile as lower 1-$\sigma$ error bar and 84\% quantile as an upper 1-$\sigma$ error bar, of linear model's slope and offset for the respective geologic unit. \label{fig:A1}}
\end{figure}

\begin{figure*}
\plotone{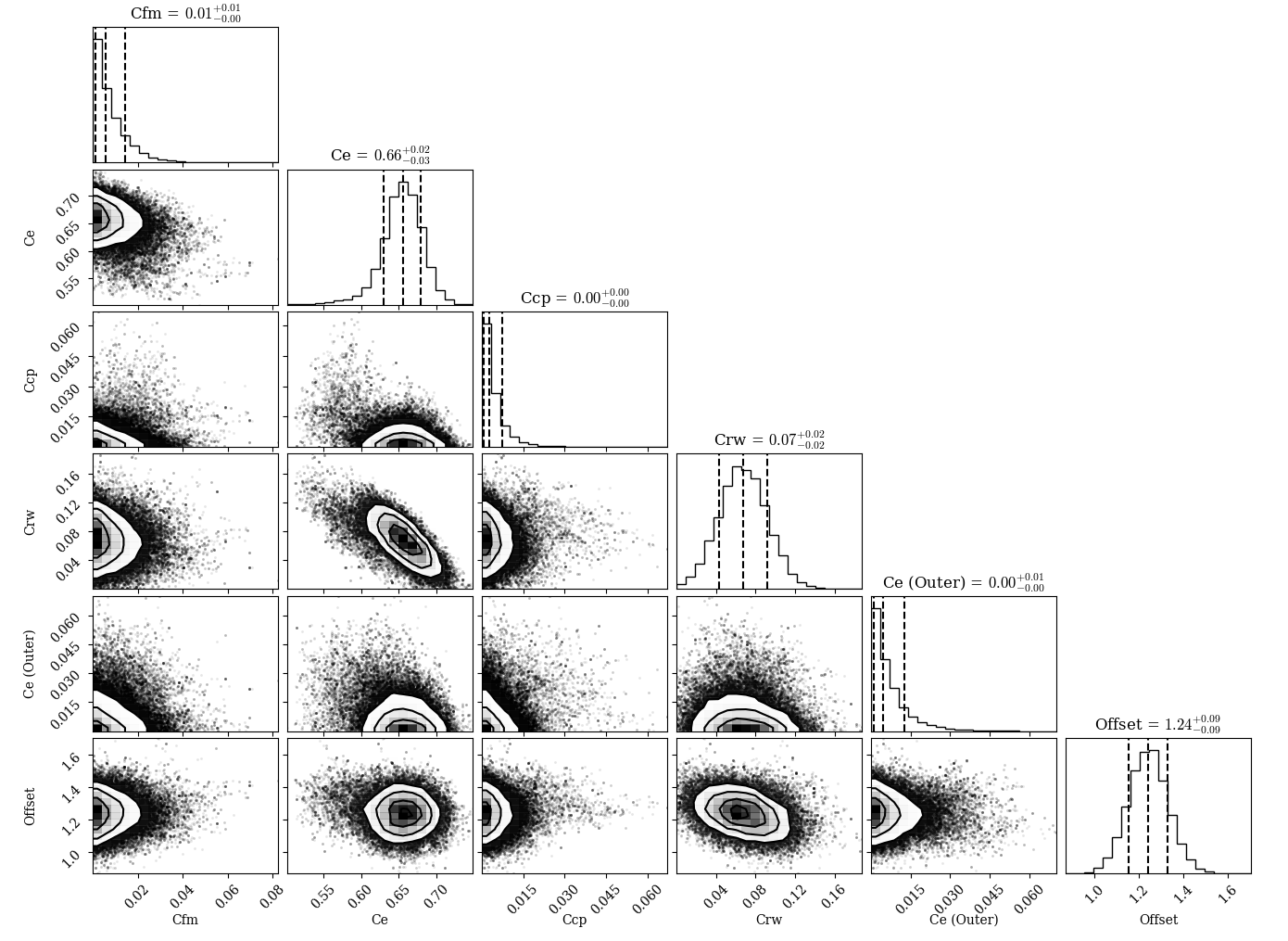}
\caption{Posterior probability distribution of the modeled parameters as the fractions contributed by each unit within the crater and outside-crater Ce (Outer) unit. The number value represents the median, including their 16\% quantile as lower 1-$\sigma$ error bar and 84\% quantile as an upper 1-$\sigma$ error bar, of fractions contributed by respective individual unit. \label{fig:A2}}
\end{figure*}

We first fit the linear model to the bulk mineralogy of the Ce (Outer) and dune field units (please see Fig. \ref{fig:A1}). The result shows a median slope value of $({0.52}^{+0.01}_{-0.01}$), the lowest likelihood unit to dune materials among the units within and beyond the crater. Using the same multiple linear regression model, we repeat the procedure of MCMC simulation using bulk mineralogy of the geologic units within the crater along with the Ce (Outer) unit. Our modeled results of the posterior probability distribution did not confirm any contribution from the Ce (Outer) unit. The median fraction contribution from Ce (Outer) unit reports no value and did not alter the median fraction contribution by geological units within the crater as reported above (please see Fig. \ref{fig:A2}). However, it is important to note that this presumption is based on the modeled results using the MCMC technique adopted here and may not be true.

Note that, even though the outside crater ejecta unit has the pyroxene and high-silica phases that are very similar to the dune field, our model is based on the overall bulk composition of the geologic units, and therefore, the result of the linear models is not determined by individual mineral endmembers (phases) alone or a combination of a few. Thus, albeit multiple mineral endmember phases have similar abundances, the overall bulk mineralogy of the geologic units determines the fractional contribution to the dune field owing to the nature of the model used. The bulk mineralogy from TES spectra shows compositional disparities between the ejecta on the floor of the crater (Ce) and the outside-crater ejecta blanket, Ce (outer) units. An interpretation of reasons of the differences between the bulk mineralogy of the ejecta on the floor of the crater and the crater's ejecta blanket is beyond the scope of the current study. A further study can be carried out to understand the reasons for these disparities by considering the mechanism of crater formation and subsequent physical and chemical weathering processes, for instance.

For reference, the list of all TES spectra used for the Ce (Outer) unit and its bulk mineralogy are reported in the appendix section of this paper (A3). We adopt this strategy as our models predict that the composition of the Ce (Outer) unit is less significant in interpreting the mineralogy of the dune field at the crater and, thus, we restrict our subsequent analyses and discussion to the geological units within the crater units only.

\begin{figure*}[ht!]
\plotone{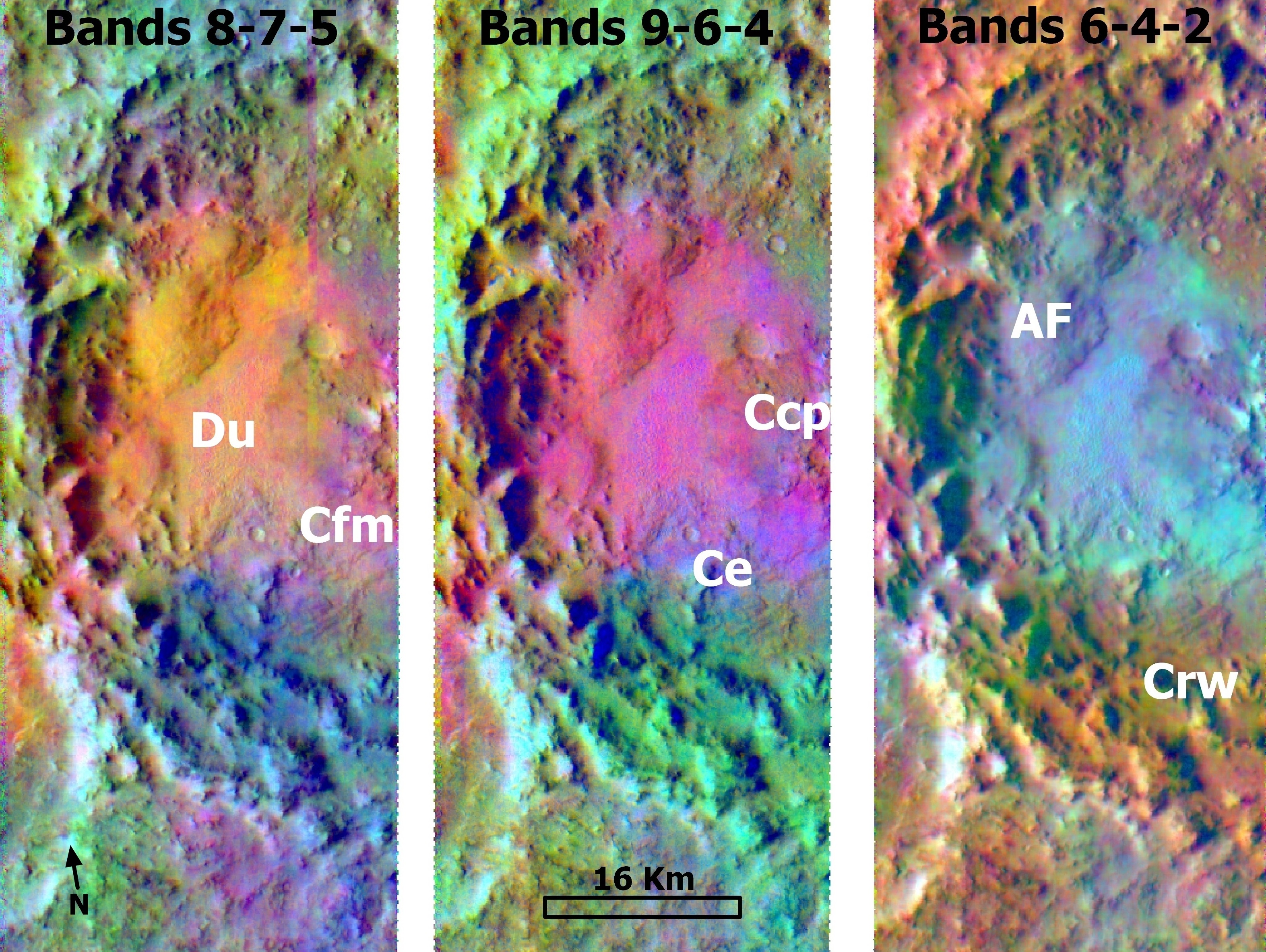}
\caption{3-panel decorrelation stretch (DCS) images using bands 8-7-5 (left), 9-6-4 (middle), and 6-4-2 (right) for the THEMIS image I02781003 overlain on the radiance image of the same scene. The geomorphic units are labeled on the DCS images: dune (Du), central peak (Ccp), floor materials (Cfm), crater ejecta (Ce), and rim and wall materials (Crw) in the color combination across all three stretches. A mixture of dark purple, bright magenta, and yellow colors in DCS 8-7-5 bands stretching across the dune field indicates a mixture of olivine, pyroxene, and Fe/Mg-smectite. The dark purple appearance on crater floor materials in DCS 8-7-5 and 9-6-4 bands is consistent with the appearance of cyan color in DCS 6-4-2 bands. The yellowish appearance (left panel) in the alluvial fan corresponds to an elevated bulk-silica content.\label{fig:09}}
\end{figure*}

The composition of the dune field was qualitatively revisited using DCS images at different band combinations, primarily of 8-7-5 bands, from THEMIS daytime IR observations (Fig. \ref{fig:09}). We used DCS image because it can be used as a proxy tool for identifying olivine-bearing basalt and silica content \citep{edwards2011mosaicking, amador2016elevated}. The dune field does not show dominance of a single color or distinctive spectral feature in DCS combinations. The 8-7-5 DCS has a mixture of dark purple, bright magenta, and yellow colors stretching across the dune field, suggesting a mixture of olivine, pyroxene, and Fe/Mg-smectite, which is a typical mineral assemblage for the Nili Fossae area (\cite{goudge2015assessing} and the references therein). The color variations of the DCS band combinations of 9-6-4 and 6-4-2 are in agreement with the interpretation of the band combination 8-7-5. For instance, the dark purple appearance on crater floor materials in DCS 8-7-5 and 9-6-4 bands is consistent with the appearance of cyan color in DCS 6-4-2 bands, representing olivine-bearing basalts \citep{edwards2011mosaicking, bandfield2011role, amador2016elevated, salvatore2016geologic}. The dune field also shows a mixture of different colors in DCS band combinations of 9-6-4 and 6-4-2, indicating a mixture of minerals.

%%%%%%%%%%%%%%%%%%%%%%%%%%%%%%%%%%%%%%%%%%%%%%%%%%%%%%%%%%%%%%%%%%%%%%%%%
\section{ Discussion} \label{sec:discuss}

The thermal inertia results suggest that the dunes are comprised of mostly medium sand-sized materials mixed with fine and coarse grain sands. Martian eolian dune fields are likely comprised of particle size of homogeneous materials within a dune field that reduces the possibility of surface mixture-induced anisothermality \citep{edwards2018thermophysical}. Thus, grain-size analysis from THEMIS thermal inertia characteristics for this dune field is considered ideal. We use the thermal inertia values across the dune field to estimate the grain size distribution. However, thermal inertia can be an incredibly ambiguous metric of physical properties of martian surface materials, and its interpretation is challenging due to several features, such as cementation, mixing of different particle sizes, presence of duricrust, ice exposure, sub-pixel-scale slope, etc. \citep{fergason2006high, piqueuxchristian2009a}. For instance, low thermal inertia ($\mathrm{<}$100~J\ m${}^{\mathrm{-}}$ ${}^{2}$\ K${}^{\mathrm{-}}$${}^{1}$\ s${}^{\mathrm{-}}$${}^{1/2}$) can result from unconsolidated dust mantling the surface of the dune field \citep{presleychristian1997b, fergason2012surface}. There are also possibilities where both vertical (layering) and lateral (horizontal mixing) mixtures \citep{presleychristian1997b, fergason2006high, piqueuxchristian2009a} could also produce a low thermal inertia value. We did not incorporate these alternative explanations for the interpretation of thermal inertia. Examination of these alternative explanations needs a future study.

The presence of potential mantling, impact cratering on the mantling surface, and thermal inertia values in the dune field can be an indication of material induration \citep{fergason2012surface}. Dunes within the crater are likely comprised of weakly indurated surfaces where there might be a consolidation that occurred in some degree to the surface and mixed with unconsolidated materials. This interpretation is consistent with the fact that a thermal inertia ($\mathrm{\sim}$200s) surface likely represents an eolian bedform overlying on an indurated regolith surface \citep{fergason2012surface}. However, the presence of thin dark streaks on the south-eastern boundary suggests a different scenario. Based on the relation between dark steaks, dunes, and subjacent terrain, the dark streaks are interpreted as the result of wind activities \citep{edgett2002low}. Gusts of wind can blow the materials out and result in streaks of deposited sands \citep{thomas1981classification, geissler2008first} toward the downwind direction (west-northwest) into the dune field. Thus, the simplest explanation of these dark streaks can be the indication of the prevalence of wind activities \citep{edgett2002low} within the crater and mobilization of the dune materials. Along the dune faces, near the crests on the western boundary, appears to be mega-ripples (i.e., TARs) spaced several meters. Bright-toned TARs in equatorial areas, for example, Syrtis Major nearby Hargraves crater, suggest an active ripple migration and presence of strong wind at the surface \citep{silvestro2020megaripple}. Thus, we hypothesize that the dune field at Hargraves is potentially active - consistent with the study of \citep{emran2020semiautomated}.

Based on the unmixing results, regression model results, and proximity of available sources of erodible materials (supported further by \ref{sec:morpho} Morphology and Stability section), we suggest that the dune materials were likely derived from physical weathering, especially eolian erosion, predominantly from the crater ejecta unit at the crater, mixed with a small amount from the crater floor and crater rim and wall lithologies. This is inferred due to their similar compositions and proximity to the dune field, situated at upwind locations in respect to the dune field. Our regression model infers a possible combination of the fraction contributions (as shown in the posterior probability distribution) from the geological units within the crater to the dune field. Albeit the similarities of bulk compositions of surrounding geological units and the dune materials gives the impression that the dunes are likely locally sourced, the use of regression models quantitatively strengthens that claim, because the regression models estimate the degree of similarities and the combination of fractional contribution by the geologic units to the dune field. However, we do not rule out the possibility that the dunes may be sourced from either one location or a varying combination of multiple provenances e.g., the mixing of two units versus input from all units within the crater. Furthermore, there is also a possibility that the dune field was sourced from only one unit that may have undergone alteration or chemical weathering that could have altered the dune field's composition.

Erosion and transport processes typically modify the relative modal abundances of constituent grains, such as the local example here where olivine is apparently enriched in the dunes. The higher concentration of olivine in the dunes versus provenance sites cannot be attributed only to the density of the olivine grains (2.9 gm/cm${}^{3}$), which is slightly less than the companion pyroxene grains (3.2 to 3.5 gm/cm${}^{3}$) \citep{lapotre2017compositional, lapotre2018morphologic}. \cite{lapotre2017compositional} suggested grain size and/or grain shape as mitigating factors in understanding compositional analysis of other martian dunes. One of these factors likely causes olivine grains to be preferentially left behind as a lag deposit, particularly on the windward side of dunes such as those in the study area and the interdune areas\citep{greeley1999aeolian, greeley2002terrestrial, jerolmack2006spatial, mangold2011segregation,hooper2012volcaniclastic}. Rather than being a pervasive lag deposit, olivine may be sequestered specifically in small coarse-grained ripples and other fine-scale bedforms of the windward sides and interdunes, which are very common in dune areas of Mars \citep{hooper2012volcaniclastic, lapotre2017compositional, lapotre2018morphologic}.

Olivine grain shape cannot be assessed directly, but the lack of cleavage in olivine and the pervasive presence of strong cleavage in pyroxenes may result in more angular and more nearly equidimensional grains as fundamental crystallographic characteristics of the olivine grain population \citep{klein2016}. Wind-drag across coarse-grained martian ripples has been suggested to cause infiltration of finer, less equidimensional grains (e.g., pyroxenes) \citep{hooper2012volcaniclastic}. These factors, combined with the slight tendency for olivine grains to be slightly larger than companion pyroxenes in Mars eolian systems \citep{lapotre2017compositional, lapotre2018morphologic} suggest that olivine may comprise a smaller amount of surficial grains in the dune-mantling ripple deposits envisioned here. However, the presumption that the olivine-pyroxene grain size and shape differences leading to pyroxene filling pore spaces is based on our modeled results, and other possible explanations may also be plausible.

Orbiter-based observations from spectroscopic measurements suggest that active dune fields and sand sheets are largely comprised of mafic minerals, such as plagioclase feldspar, high and low calcium pyroxene, and olivine \citep{achilles2017, cousin2017, ehlmann2017chemistry, lapotre2017compositional, rampe2018sand}. A strong olivine signature has also been reported elsewhere on Mars in a barchanoidal type dunes along the upwind margins of dune fields, for instance, in the Bagnold dune field as observed by high-resolution VSWIR orbital data from the Compact Reconnaissance Imaging Spectrometer for Mars \citep{seelos2014mineralogy, lapotre2017compositional, rampe2018sand}. The barchanoidal type dunes at Hargraves crater may be another instance of a dune field with elevated olivine content. We model average spectra extracted over the entire dune field (not separately from upwind and downwind margins) from a comparatively lower resolution TES footprint and, thus, we cannot confirm if there is olivine enrichment along the upwind margins. However, this may be the scenario that enriched olivine content along the upwind margins of the barchanoidal dunes at Hargraves contributes to an enhanced olivine abundance in our modeled composition.

A lower abundance of high-silica phase is reported in the dune field. Martian sands source from local to deeper craters typically contain a proportionally lower abundance of high-silica phase \citep{fenton2019bmars}. Dune fields with a lower abundance of high-silica phase are typically found on deeper crater floors whereas those with an elevated abundance ($\mathrm{>}$25\%) are found on very shallow crater floor or intercrater plains \citep{fenton2019bmars}. This statement is consistent with the geographical position of Hargraves crater and the relatively low abundance of high-silica within the dune field. Abundance of high-silica phase can be correlated with morphological characteristics such as stabilization features and bedform activity. Dune fields with elevated abundances of high-silica phase are found in southern high latitudes (60-70$\mathrm{{}^\circ}$S) where morphological characteristics of increasing stability are common \citep{fenton2019bmars}. With a stability index of 2, the dune field within Hargraves shows a relatively lower abundance of high-silica content. The yellowish appearance in DCS 8-7-5 for the alluvial fan unit represents a higher concentration of bulk-silica content \citep{amador2016elevated}. We assume that the differences in silica content between the alluvial fan and the dune field are due to the different geologic processes that are responsible in forming these features, as well as their source material. The alluvial fan is likely sourced from the crater rim or outside the crater and deposited on the crater floor through fluvial processes. 

Having a paucity of high-resolution observations from the Compact Reconnaissance Imaging Spectrometer for Mars (CRISM) \citep{murchie2007compact} across the crater floor covering all geologic units, we could not analyze reflectance spectra for the surface mineralogy. We did not use reflectance data from The Observatoire pour la Min\'{e}ralogie, l'Eau, les Glaces et l'Activit\'{e} (OMEGA) \citep{bibring2004omega} sensor because the instrument has a spatial resolution of 300 m to 5 km/pixel which is comparatively lower spatial resolution than both CRISM and THEMIS.

Bulk mineralogy in this area from TES data by \cite{salvatore2018bulk} indicates that the crater floor unit (combining the Cfm and Ccp units) has a roughly comparable phyllosilicate and amorphous component (PAC) to nearby Syrtis basalt. Their study also reported a comparable amount of plagioclase, pyroxene (both high and low calcium), and olivine to this study. There is a slight discrepancy between our result and \cite{salvatore2018bulk} when describing the bulk mineralogy of crater floor materials (for example, olivine abundance of 7$\mathrm{\pm}$5\% vs 7$\mathrm{\pm}$3\%, respectively), which is likely because they averaged spectra from the central peak and crater floor units into a single unit, as well as using different unmixing algorithms. Due to a scarcity of quality CRISM data across the crater floor, we cannot assess about finer scale individual mineral distributions for the units independently. However, the DCS image at different band combinations and the result from bulk mineralogy of TES data provide substantial information about the constituent materials of the dune field and its surrounding morphologic units.

%%%%%%%%%%%%%%%%%%%%%%%%%%%%%%%%%%%%%%%%%%%%%%%%%%%%%%%%%%%%%%%%%%%%%%%%%%%%%
\section{Conclusion}
We analyzed the thermal infrared response of the dune materials to identify the grain-size distribution and surface mineral composition of the dune field at Hargraves crater. We hypothesize that the surrounding units within the crater are the provenances for the dune field, and this paper provided modeled compositions for these units. Accordingly, we compared the dune mineralogy with the composition of surrounding geological units.

We found the average thermal inertia for the dune materials to be 238$\mathrm{\pm}$17 Jm${}^{-2}$K${}^{-1}$s${}^{-0.5}$, indicating a surface composed of an average effective grain-size of $\mathrm{\sim}$391$\mathrm{\pm}$172 $\mu$m. This size range indicates that the dune field is likely comprised of mostly medium sand-sized materials mixed with fine and coarse grain sands. The dunes are likely comprised of weakly indurated surfaces, mixed with unconsolidated materials. In this study, the determination of particle size was based on the derived thermal inertia of surface dune materials. A future study can be carried out to investigate the possibility of vertical (layering) and lateral (horizontal mixing) mixtures in the dune field.

The results from our TES compositional modeling and THEMIS DCS image analysis suggest that the dune materials are comprised of a mixture of feldspar, olivine, pyroxene, and with relatively lower bulk-silica content than the surrounding geologic units. Compositional information of the dune materials and the surrounding geologic units suggest that the dunes were likely sourced locally within the crater, predominantly from the crater ejecta, mixed with a small amount from the crater floor and crater rim and wall lithologies. The erosion and transport process likely have modified the constituent grains to some degree. We consider that the dune materials were sourced through physical weathering of the inferred provenances, based on our TES unmixing results. It is likely that through erosional and transport processes that constituent grains were modified, however, a future study is needed to investigate the possibility of alteration or chemical weathering of source materials.

Comparison of bulk mineralogy of dunes and surrounding geologic units provide useful clues on local versus distant provenances of aeolian materials. Moreover, the identification of the source provenances paves the way for a better understanding of the climate dynamics in the area of interest. For instance, the dunes at Hargraves crater are likely sourced from the geologic units within the crater- indicating the dominance of local-scale wind movement inside the crater and responsible for the deposition of sands in the dune field. It also provides insight into the material transport mechanism and erosion pattern within the crater. Since the dunes are likely locally sourced, the influence of regional and global scale dust materials is minimal in the dune field. This result coupled with the dune morphology and slipface orientation of the dunes will help in future climatic studies and geologic processes acting upon this region. On top of that, the understanding of the transport mechanism of dune materials from source to sink may provide information on recent climate changes in the region \citep{fenton2005potential}.

This study expands our knowledge in understanding the structure, composition, and eolian transport regime of the dune field at Hargraves crater through investigating grain-size, induration, modification by non-eolian processes, and dune field stability. In addition, our compositional modeling helped to constrain likely provenances and whether the dune field was sourced locally, regionally, or globally. We interpret an active and weakly indurated, simultaneously, martian dune field and showed the avenues for analyzing dunes that appear to exist on a spectrum exhibiting features of both. Our approach provides a systematic guide for interpretation and helps in expanding our understanding of both active and weakly indurated eolian environments from the observations of sands on Mars.

%%%%%%%%%%%%%%%%%%%%%%%%%%%%%%%%%%%%%%%%%%%%%%%%%%%%%%%%%%%%%%%%%%%%%%%%%%%%%%%
\begin{acknowledgments}
The authors are grateful for all of those involved in the 2001 Mars Odyssey, Mars Global Surveyor (MGS), and Mars Reconnaissance Orbiter (MRO) spacecraft missions for targeting, collecting, and archiving THEMIS, TES, CTX, and HiRISE datasets. We would like to thank Bethany Ehlmann and Amber Gullikson for providing useful information on data processing. The authors would like to thank two anonymous reviewers for their useful comments.
\end{acknowledgments}

\clearpage
%%%%%%%%%%%%%%%%%%%%%%%%%%%%%%%%%%%%%%%%%%%%%%%%%%%%%%%%%%%%%%%%%%%%%%%%%%%%%%%
\begin{appendix}

\section{Appendix information}

List of all used TES observations (A1) and spectral unmixing results of individual mineral endmembers (A2) for the geologic units within the crater, and a list of all TES spectra used for the Ce (Outer) unit and its bulk mineralogy (A3) are reported in appendix section of this paper.

\subsection{}
A list of all TES observations used in this investigation from the geologic units within the crater. For each unit, all TES observations were averaged and treated as a single thermal infrared spectrum throughout the entirety of the study (see manuscript for more information regarding the analysis of TES observations).

\startlongtable
\begin{deluxetable*}{cccc}
\tablecaption{List of all TES observations used in this investigation for the geologic units within the crater. OCK = Orbit Counter Keeper, ICK = Incremental Counter Keeper, and DET = Individual Detector.\label{tab:A1}}
\tablewidth{0pt}
\tablehead{
\colhead{Unit} & \colhead{} & \colhead{TES Observations} & \colhead{} \\
\colhead{} & \colhead{OCK} & \colhead{ICK} & \colhead{DET}
}
\decimalcolnumbers
\startdata 
Dune (Du) & 3094 & 1950 & 3 \\ 
 & 3094 & 1950 & 6 \\  
 & 3094 & 1949 & 3 \\ \hline 
\newline Crater ejecta (Ce) & 3094 & 1947 & 6 \\  
 & 3094 & 1947 & 5 \\ 
 & 3094 & 1947 & 3 \\ \hline 
\newline Crater central peak (Ccp) & 3421 & 1942 & 2 \\
 & 3421 & 1942 & 1 \\ 
 & 3421 & 1941 & 4 \\ 
 & 3421 & 1942 & 6 \\ \hline 
\newline \newline \newline Rim and wall (Crw) & 3421 & 1937 & 5 \\ 
 & 3421 & 1938 & 1 \\  
 & 3421 & 1938 & 2 \\ 
 & 3421 & 1938 & 3 \\  
 & 3094 & 1946 & 1 \\ 
 & 3421 & 1937 & 6 \\ 
 & 3094 & 1946 & 2 \\  
 & 3421 & 1938 & 6 \\ \hline 
\newline \newline Crater floor materials (Cfm) & 3421 & 1940 & 1 \\ 
 & 3421 & 1939 & 5 \\  
 & 3421 & 1940 & 2 \\ 
 & 3421 & 1940 & 4 \\ 
 & 3421 & 1940 & 5 \\ 
 & 3421 & 1939 & 6 \\ \hline 
\enddata
\end{deluxetable*}

%%%%%%%%%%%%%%%%%%%%%%%%%%%%%%%%%%%%%%%%%%%5
\subsection{}
Modeled spectral unmixing results of individual mineral endmembers for the geologic units within the crater. Reported here are the average areal abundances of endmembers (\%) and normalized for blackbody (BB) \% along with model error (RMSE).

%%%%%%%%%%%%%%%%

\subsubsection{Dune (Du) unit}

\startlongtable
\begin{deluxetable*}{ccc}
\tablecaption{Mineral endmembers for the Du unit within the crater.\label{tab:Du}}
\tablewidth{0pt}
\tablehead{
\colhead{Endmember} & \colhead{Abundance (\%)} & \colhead{Normalized for BB (\%)}
}
\decimalcolnumbers
\startdata 
Bronzite NMNH-93527 & 10.20  +/-  3.79 & 16.99 +/- 6.31 \\
Shocked An 25.5 GPa & 7.90 +/- 6.87 & 13.16 +/- 11.45 \\
Bytownite WAR-1384 & 7.43 +/- 6.95 & 12.38 +/- 11.58 \\
Swy-1 $\mathrm{<}$ 0.2 mic & 6.28 +/- 5.29 & 10.46 +/- 8.81 \\ 
KI 3362 Fo60 & 5.45 +/- 9.73 & 9.08 +/- 16.20 \\
Anorthite BUR-340 & 4.07 +/- 5.31 & 6.78 +/- 8.84 \\
Dolomite C20 & 2.99 +/- 0.52 & 4.98 +/- 0.87 \\
Avg. Lindsley pigeonite & 2.61 +/- 4.15 & 4.35 +/- 6.91 \\
Average Martian Hematite & 2.55 +/- 0.86 & 4.24 +/- 1.44 \\
Microcline BUR-3460 & 1.98 +/- 2.12 & 3.29 +/- 3.53 \\
Kieserite Kieserite & 1.84 +/- 1.11 & 3.06 +/- 1.84 \\
Crystalline heulandite (z & 1.77 +/- 5.49 & 2.94 +/- 9.14 \\  
02-011 Opal A & 1.70 +/- 1.50 & 2.83 +/- 2.50 \\ 
Quartz BUR-4120 & 0.80 +/- 1.07 & 1.34 +/- 1.78 \\
KI 3115 Fo68 & 0.75 +/- 6.68 & 1.24 +/- 11.13 \\
Fayalite WAR-RGFAY01 & 0.66 +/- 2.62 & 1.11 +/- 4.36 \\
KI 3373 Fo35 & 0.50 +/- 5.23 & 0.83 +/- 8.71 \\ 
Augite NMHN-122302 & 0.32 +/- 3.78 & 0.54 +/- 6.30 \\
Oligoclase BUR-060D & 0.15 +/- 5.11 & 0.25 +/- 8.51 \\
Enstatite HS-9.4B & 0.08 +/- 3.04 & 0.14 +/- 5.06 \\
& Sum (BB included) = 100.02 & \\ 
& Sum (BB normalized) = 100.00 & \\ 
& Blackbody Abundance = 39.99 +/- 3.21 & \\ 
& {RMS Error (\%) = 0.25} & \\
\enddata
\end{deluxetable*}

%%%%%%%%%%%%%%%%%%%%%%%%%%%%%%%%%%%
\subsubsection{Crater ejecta (Ce) unit}

\startlongtable
\begin{deluxetable*}{ccc}
\tablecaption{Mineral endmembers for the Ce unit within the crater.\label{tab:Ce}}
\tablewidth{0pt}
\tablehead{
\colhead{Endmember} & \colhead{Abundance (\%)} & \colhead{Normalized for BB (\%)}
}
\decimalcolnumbers
\startdata 
Labradorite EIBUR-3080A & 13.72 +/- 7.70 & 17.10 +/- 9.59 \\ 
Avg. Lindsley pigeonite & 12.05 +/- 3.55 & 15.02 +/- 4.43 \\ 
Bronzite NMNH-93527 & 11.28 +/- 1.85 & 14.06 +/- 2.30 \\
Bytownite WAR-1384 & 10.91 +/- 7.98 & 13.59 +/- 9.94 \\
Crystalline heulandite (z & 10.78 +/- 4.96 & 13.44 +/- 6.18 \\
Anorthite BUR-340 & 6.34 +/- 3.79 & 7.91 +/- 4.72 \\
Gypsum (Satin spar) IS6 & 3.29 +/- 1.21 & 4.10 +/- 1.50 \\
Average Martian Hematite & 3.05 +/- 0.73 & 3.81 +/- 0.91 \\ 
Dolomite C20 & 2.92 +/- 0.37 & 3.65 +/- 0.46 \\
Kieserite Kieserite & 2.39 +/- 0.81 & 2.98 +/- 1.01 \\
KI 3362 Fo60 & 1.50 +/- 2.69 & 1.87 +/- 3.35 \\
Crystalline stilbite (zeo & 0.76 +/- 2.33 & 0.94 +/- 2.90 \\
KI 3373 Fo35 & 0.62 +/- 2.86 & 0.77+/- 3.56 \\
02-011 Opal A & 0.62 +/- 0.93 & 0.77 +/- 1.16 \\
& Sum (BB included) = 99.99 & \\
& Sum (BB normalized) = 100.00 & \\
& Blackbody Abundance = 19.77 +/- 2.51 & \\ 
& {RMS Error (\%) = 0.28} & \\ 
\enddata
\end{deluxetable*}

%%%%%%%%%%%%%%%%%%%%%%%%%%%%%%%%%%%
\subsubsection{Crater central peak (Ccp) unit}

\startlongtable
\begin{deluxetable*}{ccc}
\tablecaption{Mineral endmembers for the Ccp unit within the crater.\label{tab:Ccp}}
\tablewidth{0pt}
\tablehead{
\colhead{Endmember} & \colhead{Abundance (\%)} & \colhead{Normalized for BB (\%)}
}
\decimalcolnumbers
\startdata 
Anorthite BUR-340 & 9.83 +/- 4.53 & 17.15 +/- 7.91 \\  
Bronzite NMNH-93527 & 8.94 +/- 2.22 & 15.60 +/- 3.87 \\
Crystalline heulandite (z & 7.82 +/- 7.39 & 13.63 +/- 12.90 \\
Bytownite WAR-1384 & 4.62 +/- 7.01 & 8.06 +/- 12.22 \\  
Gypsum (Satin spar) IS6 & 3.43 +/- 1.60 & 5.98 +/- 2.79 \\ 
Swy-1 $\mathrm{<}$ 0.2 mic & 3.25 +/- 5.19 & 5.67 +/- 9.05 \\
Avg. Lindsley pigeonite & 3.21 +/- 3.77 & 5.60 +/- 6.58 \\ 
Average Martian Hematite & 3.19 +/- 0.92 & 5.56 +/- 1.61 \\ 
Kieserite Kieserite & 3.09 +/- 1.08 & 5.38 +/- 1.88 \\ 
KI 3115 Fo68 & 3.00 +/- 1.74 & 5.24 +/- 3.04 \\ 
Dolomite C20 & 2.36 +/- 0.42 & 4.11 +/- 0.73 \\ 
KI 3373 Fo35 & 1.64 +/- 2.46 & 2.86 +/- 4.29 \\  
Microcline BUR-3460 & 1.05 +/- 2.34 & 1.84 +/- 4.08 \\ 
Crystalline stilbite (zeo & 0.73 +/- 3.69 & 1.28 +/- 6.43 \\ 
saponite $\mathrm{<}$0.2 mic & 0.66 +/- 1.04 & 1.14 +/- 1.81 \\ 
02-011 Opal A & 0.39 +/- 1.08 & 0.68 +/- 1.89 \\
Anhydrite S9 & 0.12 +/- 0.80 & 0.21 +/- 1.40 \\ 
& Sum (BB included) = 99.88 & \\ 
& Sum (BB normalized) = 100.00 & \\ 
& Blackbody Abundance = 42.56 +/- 2.63 & \\ 
& {RMS Error (\%) = 0.26} & \\  
\enddata
\end{deluxetable*}

%%%%%%%%%%%%%%%%%%%%%%%%%%%%%%%%%%%
\subsubsection{Crater floor material (Cfm) unit}

\startlongtable
\begin{deluxetable*}{ccc}
\tablecaption{Mineral endmembers for the Cfm unit within the crater.\label{tab:Cfm}}
\tablewidth{0pt}
\tablehead{
\colhead{Endmember} & \colhead{Abundance (\%)} & \colhead{Normalized for BB (\%)}
}
\decimalcolnumbers
\startdata 
Bytownite WAR-1384 & 11.84 +/- 4.63 & 19.38 +/- 7.58 \\ 
Bronzite NMNH-93527 & 11.23 +/- 1.84 & 18.39 +/- 3.02 \\
Crystalline heulandite (z & 8.71 +/- 4.12 & 14.27 +/- 6.75 \\ 
Augite NMHN-122302 & 3.75 +/- 2.64 & 6.15 +/- 4.32 \\  
Kieserite Kieserite & 3.35 +/- 0.66 & 5.48 +/- 1.09 \\  
KI 3115 Fo68 & 3.21 +/- 1.55 & 5.25 +/- 2.54 \\  
Swy-1 $\mathrm{<}$ 0.2 mic & 3.09 +/- 4.16 & 5.06 +/- 6.81 \\ 
Anorthite BUR-340 & 3.02 +/- 3.83 & 4.95 +/- 6.26 \\  
Avg. Lindsley pigeonite & 3.02 +/- 3.36 & 4.94 +/- 5.51 \\ 
Gypsum (Satin spar) IS6 & 2.69 +/- 0.97 & 4.40 +/- 1.59 \\ 
Average Martian Hematite & 2.28 +/- 0.88 & 3.74 +/- 1.44 \\ 
Crystalline stilbite (zeo & 1.67 +/- 2.45 & 2.74 +/- 4.00 \\
Dolomite C20 & 1.56 +/- 0.40 & 2.55 +/- 0.66 \\ 
Fayalite WAR-RGFAY01 & 0.92 +/- 4.25 & 1.51 +/- 6.97 \\ 
K-rich Glass & 0.40 +/- 2.24 & 0.65 +/- 3.66 \\
KI 3373 Fo35 & 0.25 +/- 3.64 & 0.41 +/- 5.97 \\
KI 3008 Fo10 & 0.10 +/- 4.52 & 0.16 +/- 7.40 \\ 
& Sum (BB included) = 100.00 & \\  
& Sum (BB normalized) = 100.00 & \\ 
& Blackbody Abundance = 38.92 +/- 2.22 &  \\ 
& {RMS Error (\%) = 0.22} & \\
\enddata
\end{deluxetable*}

%%%%%%%%%%%%%%%%%%%%%%%%%%%%%%%%%%%
\subsubsection{Crater rim and wall (Crw) unit}

\startlongtable
\begin{deluxetable*}{ccc}
\tablecaption{Mineral endmembers for the Crw unit within the crater.\label{tab:Crw}}
\tablewidth{0pt}
\tablehead{
\colhead{Endmember} & \colhead{Abundance (\%)} & \colhead{Normalized for BB (\%)}
}
\decimalcolnumbers
\startdata 
Bytownite WAR-1384 & 10.21 +/- 4.85 & 22.27 +/- 10.58 \\  
Crystalline heulandite (z & 6.54 +/- 4.10 & 14.26 +/- 8.93 \\ 
Shocked An 21 GPa & 4.97 +/- 4.42 & 10.85 +/- 9.64 \\ 
Bronzite NMNH-93527 & 4.79 +/- 2.66 & 10.45 +/- 5.81 \\  
Avg. Lindsley pigeonite & 4.33 +/- 2.34 & 9.45 +/- 5.11 \\  
Swy-1 $\mathrm{<}$ 0.2 mic & 3.32 +/- 3.81 & 7.23 +/- 8.30 \\  
Average Martian Hematite & 2.72+/- 0.60 & 5.94 +/- 1.31 \\ 
Kieserite Kieserite & 2.27 +/- 0.71 & 4.95 +/- 1.55 \\ 
Gypsum (Satin spar) IS6 & 1.68 +/- 1.01 & 3.67 +/- 2.20 \\ 
Dolomite C20 & 1.40 +/- 0.28 & 3.05 +/- 0.61 \\ 
Anorthite BUR-340 & 1.12 +/- 3.91 & 2.43 +/- 8.52 \\
KI 3362 Fo60 & 0.91 +/- 3.39 & 1.99 +/- 7.40 \\
Quartz BUR-4120 & 0.50 +/- 0.45 & 1.08 +/- 0.98 \\ 
KI 3115 Fo68 & 0.47 +/- 2.78 & 1.03 +/- 6.07 \\ 
Microcline BUR-3460 & 0.30 +/- 1.59 & 0.65+/- 3.47 \\ 
Enstatite HS-9.4B & 0.29 +/- 1.94 & 0.62 +/- 4.24 \\ 
Albite WAR-0235 & 0.02 +/- 1.38 & 0.04 +/- 3.01 \\ 
Anhydrite S9 & 0.01 +/- 0.54 & 0.03 +/- 1.17 \\ 
& Sum (BB included) = 99.97 & \\ 
& Sum (BB normalized) = 100.00 & \\  
& Blackbody Abundance = 54.12 +/- 2.32 & \\
& {RMS Error (\%) = 0.17} &\\ 
\enddata
\end{deluxetable*}

%%%%%%%%%%%%%%%%%%%%%%%%%%%%%%%%%%%%%%%
\subsection{}
\subsubsection{}
A list of all TES observations used in this investigation for the Ce (Outer) unit. All TES observations were averaged and treated as a single thermal infrared spectrum throughout the entirety of the study (see manuscript for more information regarding the analysis of TES observations).

\startlongtable
\begin{deluxetable*}{cccccccccc}
\tablecaption{List of all TES observations used in this investigation for the Ce (Outer) unit. OCK = Orbit Counter Keeper, ICK = Incremental Counter Keeper, and DET = Individual Detector.}
\tablewidth{0pt}
\tablehead{
\colhead{Unit} & \colhead{}  & \colhead{}  & \colhead{}  & \colhead{TES} & \colhead{Observations} & \colhead{} & \colhead{} & \colhead{} & \colhead{} \\
\colhead{} & \colhead{OCK} & \colhead{ICK} & \colhead{DET}  & \colhead{OCK} & \colhead{ICK} & \colhead{DET}  & \colhead{OCK} & \colhead{ICK} & \colhead{DET}
}
\decimalcolnumbers
\startdata 
 & 3836 & 1944 & 1 & 3836 & 1944 & 4 & 3836 & 1943 & 5 \\  
 & 3836 & 1944 & 3 & 3836 & 1942 & 4 & 3836 & 1945 & 4 \\ 
 & 3836 & 1942 & 3 & 3836 & 1944 & 6 & 3836 & 1945 & 6 \\ 
 Ce (Outer) & 3836 & 1942 & 2 & 3836 & 1942 & 6 & 3836 & 1948 & 1 \\  
 & 3836 & 1944 & 5 & 3836 & 1945 & 1 & 3836 & 1947 & 1 \\
 & 3836 & 1942 & 5 & 3836 & 1945 & 3 & 3836 & 1949 & 1 \\ 
 & 3836 & 1949 & 2 & 3836 & 1949 & 5 & 3836 & 1948 & 5 \\ 
 & 3836 & 1948 & 4 & 3836 & 1948 & 6 & 3836 & 1949 & 3 \\ 
\enddata
\end{deluxetable*}

\subsubsection{}
\startlongtable
\begin{deluxetable*}{ccccccccccc}
\tablecaption{Our spectral unmixing result for Ce (outer) unit. Reported here are the average areal abundances of each mineral group (\%) along with the calculated model error (RMSE). All values were normalized to the blackbody (BB) percentage (\%).}
\tablewidth{0pt}
\tablehead{
\colhead{Groups} & \colhead{Feldspar}  & \colhead{Pyroxene}  & \colhead{Olivine}  & \colhead{HSP} & \colhead{Carbonate} & \colhead{Sulfate} & \colhead{Hematite} & \colhead{Quartz} & \colhead{BB} & \colhead{RMSE} 
}
\decimalcolnumbers
\startdata 
\% & 28$\mathrm{\pm}$4 & 25$\mathrm{\pm}$4 & 8$\mathrm{\pm}$2 & 17$\mathrm{\pm}$6 & 3$\mathrm{\pm}$1 & 9$\mathrm{\pm}$2 & 8$\mathrm{\pm}$2 & 2$\mathrm{\pm}$1 & 0.26$\mathrm{\pm}$3 & 0.21 \\ 
\enddata
\tablecomments{HSP = High Silica Phase}
\end{deluxetable*}

\end{appendix}

\bibliography{sample631}{}

\begin{thebibliography}{}
\expandafter\ifx\csname natexlab\endcsname\relax\def\natexlab#1{#1}\fi
\providecommand{\url}[1]{\href{#1}{#1}}
\providecommand{\dodoi}[1]{doi:~\href{http://doi.org/#1}{\nolinkurl{#1}}}
\providecommand{\doeprint}[1]{\href{http://ascl.net/#1}{\nolinkurl{http://ascl.net/#1}}}
\providecommand{\doarXiv}[1]{\href{https://arxiv.org/abs/#1}{\nolinkurl{https://arxiv.org/abs/#1}}}

\bibitem[{Abuodha(2003)}]{abuodha2003grain}
Abuodha, J. 2003, Journal of African Earth Sciences, 36, 41

\bibitem[{Achilles {et~al.}(2017)Achilles, Downs, Ming, Rampe, Morris, Treiman,
  Morrison, Blake, Vaniman, Ewing, {et~al.}}]{achilles2017}
Achilles, C., Downs, R., Ming, D., {et~al.} 2017, Journal of Geophysical
  Research: Planets, 122, 2344

\bibitem[{Amador \& Bandfield(2016)}]{amador2016elevated}
Amador, E.~S., \& Bandfield, J.~L. 2016, Icarus, 276, 39

\bibitem[{Baldridge(2008)}]{baldridge2008thermal}
Baldridge, A. 2008, PhD thesis, Ph. D. Dissertation thesis, 204 pp., Arizona
  State University, Tempe, AZ

\bibitem[{Bandeira {et~al.}(2010)Bandeira, Marques, Saraiva, \&
  Pina}]{bandeira2010automated}
Bandeira, L., Marques, J.~S., Saraiva, J., \& Pina, P. 2010, in International
  Conference Image Analysis and Recognition, Springer, 306--315

\bibitem[{Bandfield(2002)}]{bandfield2002global}
Bandfield, J.~L. 2002, Journal of Geophysical Research: Planets, 107, 9

\bibitem[{Bandfield {et~al.}(2000)Bandfield, Christensen, \&
  Smith}]{bandfield2000spectral}
Bandfield, J.~L., Christensen, P.~R., \& Smith, M.~D. 2000, Journal of
  Geophysical Research: Planets, 105, 9573

\bibitem[{Bandfield {et~al.}(2004{\natexlab{a}})Bandfield, Hamilton,
  Christensen, \& McSween~Jr}]{bandfield2004a}
Bandfield, J.~L., Hamilton, V.~E., Christensen, P.~R., \& McSween~Jr, H.~Y.
  2004{\natexlab{a}}, Journal of Geophysical Research: Planets, 109

\bibitem[{Bandfield {et~al.}(2011)Bandfield, Rogers, \&
  Edwards}]{bandfield2011role}
Bandfield, J.~L., Rogers, A.~D., \& Edwards, C.~S. 2011, Icarus, 211, 157

\bibitem[{Bandfield {et~al.}(2004{\natexlab{b}})Bandfield, Rogers, Smith, \&
  Christensen}]{bandfield2004b}
Bandfield, J.~L., Rogers, D., Smith, M.~D., \& Christensen, P.~R.
  2004{\natexlab{b}}, Journal of Geophysical Research: Planets, 109

\bibitem[{Banks {et~al.}(2018)Banks, Fenton, Bridges, Geissler, Chojnacki,
  Runyon, Silvestro, \& Zimbelman}]{banks2018patterns}
Banks, M.~E., Fenton, L.~K., Bridges, N.~T., {et~al.} 2018, Journal of
  Geophysical Research: Planets, 123, 3205

\bibitem[{Bibring {et~al.}(2004)Bibring, Soufflot, Berth{\'e}, Langevin,
  Gondet, Drossart, Bouy{\'e}, Combes, Puget, Semery,
  {et~al.}}]{bibring2004omega}
Bibring, J.-P., Soufflot, A., Berth{\'e}, M., {et~al.} 2004, in Mars Express:
  the scientific payload, Vol. 1240, 37--49

\bibitem[{Charles {et~al.}(2017)Charles, Titus, Hayward, Edwards, \&
  Ahrens}]{charles2017comparison}
Charles, H., Titus, T., Hayward, R., Edwards, C., \& Ahrens, C. 2017, Earth and
  Planetary Science Letters, 458, 152

\bibitem[{Christensen {et~al.}(2009)Christensen, Engle, Anwar, Dickenshied,
  Noss, Gorelick, \& Weiss-Malik}]{christensen2009jmars}
Christensen, P., Engle, E., Anwar, S., {et~al.} 2009, in AGU Fall Meeting
  Abstracts, Vol. 2009, IN22A--06

\bibitem[{Christensen {et~al.}(2013)Christensen, Fergason, Edwards, \&
  Hill}]{christensen2013themis}
Christensen, P., Fergason, R., Edwards, C., \& Hill, J. 2013, in Lunar and
  Planetary Science Conference No. 1719, 2822

\bibitem[{Christensen {et~al.}(2000)Christensen, Bandfield, Hamilton, Howard,
  Lane, Piatek, Ruff, \& Stefanov}]{christensen2000thermal}
Christensen, P.~R., Bandfield, J.~L., Hamilton, V.~E., {et~al.} 2000, Journal
  of Geophysical Research: Planets, 105, 9735

\bibitem[{Christensen {et~al.}(2001)Christensen, Bandfield, Hamilton, Ruff,
  Kieffer, Titus, Malin, Morris, Lane, Clark, {et~al.}}]{christensen2001mars}
---. 2001, Journal of Geophysical Research: Planets, 106, 23823

\bibitem[{Christensen {et~al.}(2004)Christensen, Jakosky, Kieffer, Malin,
  McSween, Nealson, Mehall, Silverman, Ferry, Caplinger,
  {et~al.}}]{christensen2004thermal}
Christensen, P.~R., Jakosky, B.~M., Kieffer, H.~H., {et~al.} 2004, Space
  Science Reviews, 110, 85

\bibitem[{Courville {et~al.}(2016)Courville, Putzig, Hoover, \&
  Fenton}]{courville2016thermophysical}
Courville, S.~W., Putzig, N.~E., Hoover, R., \& Fenton, L.~K. 2016, in AGU Fall
  Meeting Abstracts, P21A--2073

\bibitem[{Cousin {et~al.}(2017)Cousin, Dehouck, Meslin, Forni, Williams, Stein,
  Gasnault, Bridges, Ehlmann, Schr{\"o}der, {et~al.}}]{cousin2017}
Cousin, A., Dehouck, E., Meslin, P.-Y., {et~al.} 2017, Journal of Geophysical
  Research: Planets, 122, 2144

\bibitem[{Edgett(2002)}]{edgett2002low}
Edgett, K.~S. 2002, Journal of Geophysical Research: Planets, 107, 5

\bibitem[{Edgett \& Christensen(1991)}]{edgett1991particle}
Edgett, K.~S., \& Christensen, P.~R. 1991, Journal of Geophysical Research:
  Planets, 96, 22765

\bibitem[{Edgett \& Malin(2000)}]{edgett2000new}
Edgett, K.~S., \& Malin, M.~C. 2000, Journal of Geophysical Research: Planets,
  105, 1623

\bibitem[{Edwards {et~al.}(2015)Edwards, Anwar, Hagee, Doerres, Dickensheid, \&
  Christensen}]{edwards2015processing}
Edwards, C., Anwar, S., Hagee, W., {et~al.} 2015, in Second Planetary Data
  Workshop, Vol. 1846, 7032

\bibitem[{Edwards {et~al.}(2009)Edwards, Bandfield, Christensen, \&
  Fergason}]{edwards2009global}
Edwards, C., Bandfield, J., Christensen, P., \& Fergason, R. 2009, Journal of
  Geophysical Research: Planets, 114

\bibitem[{Edwards {et~al.}(2011)Edwards, Nowicki, Christensen, Hill, Gorelick,
  \& Murray}]{edwards2011mosaicking}
Edwards, C., Nowicki, K., Christensen, P., {et~al.} 2011, Journal of
  Geophysical Research: Planets, 116

\bibitem[{Edwards {et~al.}(2018)Edwards, Piqueux, Hamilton, Fergason,
  Herkenhoff, Vasavada, Bennett, Sacks, Lewis, \&
  Smith}]{edwards2018thermophysical}
Edwards, C.~S., Piqueux, S., Hamilton, V.~E., {et~al.} 2018, Journal of
  Geophysical Research: Planets, 123, 1307

\bibitem[{Ehlmann {et~al.}(2017)Ehlmann, Edgett, Sutter, Achilles, Litvak,
  Lapotre, Sullivan, Fraeman, Arvidson, Blake, {et~al.}}]{ehlmann2017chemistry}
Ehlmann, B., Edgett, K., Sutter, B., {et~al.} 2017, Journal of Geophysical
  Research: Planets, 122, 2510

\bibitem[{Emran(2019)}]{emran2019surficial}
Emran, A. 2019

\bibitem[{Emran {et~al.}(2019{\natexlab{a}})Emran, Marzen, \&
  King}]{emran2019b}
Emran, A., Marzen, L., \& King, D. 2019{\natexlab{a}}, 3,
  \dodoi{10.17632/bfnxzwhd4r.3}

\bibitem[{Emran {et~al.}(2019{\natexlab{b}})Emran, Marzen, \&
  King~Jr}]{emran2019a}
Emran, A., Marzen, L., \& King~Jr, D. 2019{\natexlab{b}}

\bibitem[{Emran {et~al.}(2020)Emran, Marzen, \&
  King~Jr}]{emran2020semiautomated}
---. 2020, Earth and Space Science, 7, e2019EA000935

\bibitem[{Feely \& Christensen(1999)}]{feelychristian1999}
Feely, K.~C., \& Christensen, P.~R. 1999, Journal of Geophysical Research:
  Planets, 104, 24195

\bibitem[{Feldman {et~al.}(2004)Feldman, Prettyman, Maurice, Plaut, Bish,
  Vaniman, Mellon, Metzger, Squyres, Karunatillake,
  {et~al.}}]{feldman2004global}
Feldman, W., Prettyman, T., Maurice, S., {et~al.} 2004, Journal of Geophysical
  Research: Planets, 109

\bibitem[{Fenton {et~al.}(2019{\natexlab{a}})Fenton, Gullikson, Hayward,
  Charles, \& Titus}]{fenton2019amars}
Fenton, L., Gullikson, A., Hayward, R., Charles, H., \& Titus, T.
  2019{\natexlab{a}}, in Lunar and Planetary Science Conference No. 2132, 1115

\bibitem[{Fenton(2005)}]{fenton2005potential}
Fenton, L.~K. 2005, Journal of Geophysical Research: Planets, 110

\bibitem[{Fenton {et~al.}(2019{\natexlab{b}})Fenton, Gullikson, Hayward,
  Charles, \& Titus}]{fenton2019bmars}
Fenton, L.~K., Gullikson, A.~L., Hayward, R.~K., Charles, H., \& Titus, T.~N.
  2019{\natexlab{b}}, Icarus, 330, 189

\bibitem[{Fenton \& Hayward(2010)}]{fentonhayward2010southern}
Fenton, L.~K., \& Hayward, R.~K. 2010, Geomorphology, 121, 98

\bibitem[{Fenton \& Mellon(2006)}]{fenton2006thermal}
Fenton, L.~K., \& Mellon, M.~T. 2006, Journal of Geophysical Research: Planets,
  111

\bibitem[{Fenton {et~al.}(2005)Fenton, Toigo, \&
  Richardson}]{fenton2005aeolian}
Fenton, L.~K., Toigo, A.~D., \& Richardson, M.~I. 2005, Journal of Geophysical
  Research: Planets, 110

\bibitem[{Fergason {et~al.}(2012)Fergason, Christensen, Golombek, \&
  Parker}]{fergason2012surface}
Fergason, R., Christensen, P., Golombek, M., \& Parker, T. 2012, Space science
  reviews, 170, 739

\bibitem[{Fergason {et~al.}(2006)Fergason, Christensen, \&
  Kieffer}]{fergason2006high}
Fergason, R.~L., Christensen, P.~R., \& Kieffer, H.~H. 2006, Journal of
  Geophysical Research: Planets, 111

\bibitem[{Foreman-Mackey {et~al.}(2013)Foreman-Mackey, Hogg, Lang, \&
  Goodman}]{foreman2013emcee}
Foreman-Mackey, D., Hogg, D.~W., Lang, D., \& Goodman, J. 2013, Publications of
  the Astronomical Society of the Pacific, 125, 306

\bibitem[{Gardin {et~al.}(2012)Gardin, Allemand, Quantin, Silvestro, \&
  Delacourt}]{gardin2012dune}
Gardin, E., Allemand, P., Quantin, C., Silvestro, S., \& Delacourt, C. 2012,
  Planetary and Space Science, 60, 314

\bibitem[{Geissler {et~al.}(2008)Geissler, Johnson, Sullivan, Herkenhoff,
  Mittlefehldt, Fergason, Ming, Morris, Squyres, Soderblom,
  {et~al.}}]{geissler2008first}
Geissler, P.~E., Johnson, J., Sullivan, R., {et~al.} 2008, Journal of
  Geophysical Research: Planets, 113

\bibitem[{Gillespie {et~al.}(1986)Gillespie, Kahle, \&
  Walker}]{gillespie1986color}
Gillespie, A.~R., Kahle, A.~B., \& Walker, R.~E. 1986, Remote Sensing of
  Environment, 20, 209

\bibitem[{Glotch {et~al.}(2004)Glotch, Morris, Christensen, \&
  Sharp}]{glotch2004effect}
Glotch, T., Morris, R., Christensen, P., \& Sharp, T. 2004, Journal of
  Geophysical Research: Planets, 109

\bibitem[{Goodman \& Weare(2010)}]{goodmanweare2010}
Goodman, J., \& Weare, J. 2010, Communications in applied mathematics and
  computational science, 5, 65

\bibitem[{Goudge {et~al.}(2015)Goudge, Mustard, Head, Fassett, \&
  Wiseman}]{goudge2015assessing}
Goudge, T.~A., Mustard, J.~F., Head, J.~W., Fassett, C.~I., \& Wiseman, S.~M.
  2015, Journal of Geophysical Research: Planets, 120, 775

\bibitem[{Greeley {et~al.}(2002)Greeley, Bridges, Kuzmin, \&
  Laity}]{greeley2002terrestrial}
Greeley, R., Bridges, N.~T., Kuzmin, R.~O., \& Laity, J.~E. 2002, Journal of
  Geophysical Research: Planets, 107, 5

\bibitem[{Greeley \& Iversen(1987)}]{greeley1987wind}
Greeley, R., \& Iversen, J.~D. 1987, Wind as a geological process: on Earth,
  Mars, Venus and Titan No.~4 (CUP Archive)

\bibitem[{Greeley {et~al.}(2001)Greeley, Kuzmin, \&
  Haberle}]{greeley2001aeolian}
Greeley, R., Kuzmin, R.~O., \& Haberle, R.~M. 2001, Space Science Reviews, 96,
  393

\bibitem[{Greeley {et~al.}(1999)Greeley, Kraft, Sullivan, Wilson, Bridges,
  Herkenhoff, Kuzmin, Malin, \& Ward}]{greeley1999aeolian}
Greeley, R., Kraft, M., Sullivan, R., {et~al.} 1999, Journal of Geophysical
  Research: Planets, 104, 8573

\bibitem[{Gullikson {et~al.}(2018)Gullikson, Hayward, Titus, Charles, Fenton,
  Hoover, \& Putzig}]{gullikson2018mars}
Gullikson, A.~L., Hayward, R.~K., Titus, T.~N., {et~al.} 2018, US Geological
  Survey Open-File Report, 1164

\bibitem[{Haberle \& Jakosky(1991)}]{haberle1991atmospheric}
Haberle, R.~M., \& Jakosky, B.~M. 1991, Icarus, 90, 187

\bibitem[{Hayward {et~al.}(2010)Hayward, Fenton, Tanaka, Titus, Colaprete, \&
  Christensen}]{hayward2010mars}
Hayward, R., Fenton, L., Tanaka, K., {et~al.} 2010, US Geological Survey:
  Flagstaff, AZ, USA

\bibitem[{Hayward {et~al.}(2012)Hayward, Fenton, Titus, Colaprete, \&
  Christensen}]{hayward2012mars}
Hayward, R., Fenton, L., Titus, T., Colaprete, A., \& Christensen, P. 2012, US
  Geological Survey: Flagstaff, AZ, USA

\bibitem[{Hayward {et~al.}(2007{\natexlab{a}})Hayward, Mullins, Fenton, Hare,
  Titus, Bourke, Colaprete, \& Christensen}]{hayward2007amars}
Hayward, R., Mullins, K., Fenton, L., {et~al.} 2007{\natexlab{a}}

\bibitem[{Hayward {et~al.}(2014)Hayward, Fenton, \& Titus}]{hayward2014mars}
Hayward, R.~K., Fenton, L., \& Titus, T.~N. 2014, Icarus, 230, 38

\bibitem[{Hayward {et~al.}(2007{\natexlab{b}})Hayward, Mullins, Fenton, Hare,
  Titus, Bourke, Colaprete, \& Christensen}]{hayward2007bmars}
Hayward, R.~K., Mullins, K.~F., Fenton, L.~K., {et~al.} 2007{\natexlab{b}},
  Journal of Geophysical Research: Planets, 112

\bibitem[{Hayward {et~al.}(2009)Hayward, Titus, Michaels, Fenton, Colaprete, \&
  Christensen}]{hayward2009aeolian}
Hayward, R.~K., Titus, T.~N., Michaels, T.~I., {et~al.} 2009, Journal of
  Geophysical Research: Planets, 114

\bibitem[{Hiesinger \& Head~III(2004)}]{hiesingerhead2004syrtis}
Hiesinger, H., \& Head~III, J. 2004, Journal of Geophysical Research: Planets,
  109

\bibitem[{Hogg \& Foreman-Mackey(2018)}]{hogg2018data}
Hogg, D.~W., \& Foreman-Mackey, D. 2018, The Astrophysical Journal Supplement
  Series, 236, 11

\bibitem[{Hooper {et~al.}(2012)Hooper, McGinnis, \&
  Necsoiu}]{hooper2012volcaniclastic}
Hooper, D.~M., McGinnis, R.~N., \& Necsoiu, M. 2012, Earth Surface Processes
  and Landforms, 37, 1090

\bibitem[{Ivanov {et~al.}(2012)Ivanov, Hiesinger, Erkeling, Hielscher, \&
  Reiss}]{ivanov2012major}
Ivanov, M., Hiesinger, H., Erkeling, G., Hielscher, F., \& Reiss, D. 2012,
  Icarus, 218, 24

\bibitem[{Jakosky {et~al.}(2000)Jakosky, Mellon, Kieffer, Christensen, Varnes,
  \& Lee}]{jakosky2000thermal}
Jakosky, B.~M., Mellon, M.~T., Kieffer, H.~H., {et~al.} 2000, Journal of
  Geophysical Research: Planets, 105, 9643

\bibitem[{Jerolmack {et~al.}(2006)Jerolmack, Mohrig, Grotzinger, Fike, \&
  Watters}]{jerolmack2006spatial}
Jerolmack, D.~J., Mohrig, D., Grotzinger, J.~P., Fike, D.~A., \& Watters, W.~A.
  2006, Journal of Geophysical Research: Planets, 111

\bibitem[{Johnson {et~al.}(2002)Johnson, H{\"o}rz, Lucey, \&
  Christensen}]{johnson2002thermal}
Johnson, J.~R., H{\"o}rz, F., Lucey, P.~G., \& Christensen, P.~R. 2002, Journal
  of Geophysical Research: Planets, 107, 3

\bibitem[{Kieffer {et~al.}(1973)Kieffer, Chase~Jr, Miner, M{\"u}nch, \&
  Neugebauer}]{kieffer1973preliminary}
Kieffer, H., Chase~Jr, S., Miner, E., M{\"u}nch, G., \& Neugebauer, G. 1973,
  Journal of Geophysical Research, 78, 4291

\bibitem[{Kieffer {et~al.}(1977)Kieffer, Martin, Peterfreund, Jakosky, Miner,
  \& Palluconi}]{kieffer1977thermal}
Kieffer, H.~H., Martin, T., Peterfreund, A.~R., {et~al.} 1977, Journal of
  Geophysical Research, 82, 4249

\bibitem[{Klein \& Philpotts(2016)}]{klein2016}
Klein, C., \& Philpotts, A. 2016, Earth {{Materials}} 2nd {{Edition}}:
  {{Introduction}} to {{Mineralogy}} and {{Petrology}} ({Cambridge University
  Press})

\bibitem[{Koeppen \& Hamilton(2008)}]{koeppenhamilton2008global}
Koeppen, W.~C., \& Hamilton, V.~E. 2008, Journal of Geophysical Research:
  Planets, 113

\bibitem[{Lapotre {et~al.}(2018)Lapotre, Ewing, Weitz, Lewis, Lamb, Ehlmann, \&
  Rubin}]{lapotre2018morphologic}
Lapotre, M., Ewing, R., Weitz, C., {et~al.} 2018, Geophysical Research Letters,
  45, 10

\bibitem[{Lapotre {et~al.}(2017)Lapotre, Ehlmann, Minson, Arvidson, Ayoub,
  Fraeman, Ewing, \& Bridges}]{lapotre2017compositional}
Lapotre, M.~G., Ehlmann, B., Minson, S.~E., {et~al.} 2017, Journal of
  Geophysical Research: Planets, 122, 2489

\bibitem[{Malin {et~al.}(2007)Malin, Bell, Cantor, Caplinger, Calvin, Clancy,
  Edgett, Edwards, Haberle, James, {et~al.}}]{malin2007context}
Malin, M.~C., Bell, J.~F., Cantor, B.~A., {et~al.} 2007, Journal of Geophysical
  Research: Planets, 112

\bibitem[{Mangold {et~al.}(2011)Mangold, Baratoux, Arnalds, Bardintzeff,
  Platevoet, Gr{\'e}goire, \& Pinet}]{mangold2011segregation}
Mangold, N., Baratoux, D., Arnalds, O., {et~al.} 2011, Earth and Planetary
  Science Letters, 310, 233

\bibitem[{Mangold {et~al.}(2007)Mangold, Poulet, Mustard, Bibring, Gondet,
  Langevin, Ansan, Masson, Fassett, Head~III, {et~al.}}]{mangold2007mineralogy}
Mangold, N., Poulet, F., Mustard, J., {et~al.} 2007, Journal of Geophysical
  Research: Planets, 112

\bibitem[{McEwen {et~al.}(2007)McEwen, Eliason, Bergstrom, Bridges, Hansen,
  Delamere, Grant, Gulick, Herkenhoff, Keszthelyi, {et~al.}}]{mcewen2007mars}
McEwen, A.~S., Eliason, E.~M., Bergstrom, J.~W., {et~al.} 2007, Journal of
  Geophysical Research: Planets, 112

\bibitem[{McKee(1979)}]{mckee1979study}
McKee, E.~D. 1979, A study of global sand seas, Vol. 1052 (US Government
  Printing Office)

\bibitem[{Mellon {et~al.}(2000)Mellon, Jakosky, Kieffer, \&
  Christensen}]{mellon2000high}
Mellon, M.~T., Jakosky, B.~M., Kieffer, H.~H., \& Christensen, P.~R. 2000,
  Icarus, 148, 437

\bibitem[{Michalski {et~al.}(2003)Michalski, Kraft, Diedrich, Sharp, \&
  Christensen}]{michalski2003thermal}
Michalski, J.~R., Kraft, M.~D., Diedrich, T., Sharp, T.~G., \& Christensen,
  P.~R. 2003, Geophysical Research Letters, 30

\bibitem[{Michalski {et~al.}(2005)Michalski, Kraft, Sharp, Williams, \&
  Christensen}]{michalski2005mineralogical}
Michalski, J.~R., Kraft, M.~D., Sharp, T.~G., Williams, L.~B., \& Christensen,
  P.~R. 2005, Icarus, 174, 161

\bibitem[{Michalski {et~al.}(2006)Michalski, Kraft, Sharp, Williams, \&
  Christensen}]{michalski2006emission}
---. 2006, Journal of Geophysical Research: Planets, 111

\bibitem[{Murchie {et~al.}(2007)Murchie, Arvidson, Bedini, Beisser, Bibring,
  Bishop, Boldt, Cavender, Choo, Clancy, {et~al.}}]{murchie2007compact}
Murchie, S., Arvidson, R., Bedini, P., {et~al.} 2007, Journal of Geophysical
  Research: Planets, 112

\bibitem[{Neugebauer {et~al.}(1971)Neugebauer, M{\"u}nch, Kieffer, Chase~Jr, \&
  Miner}]{neugebauer1971mariner}
Neugebauer, G., M{\"u}nch, G., Kieffer, H., Chase~Jr, S., \& Miner, E. 1971,
  Astronomical Journal, 76, 719

\bibitem[{Piqueux \& Christensen(2009{\natexlab{a}})}]{piqueuxchristian2009a}
Piqueux, S., \& Christensen, P. 2009{\natexlab{a}}, Journal of Geophysical
  Research: Planets, 114

\bibitem[{Piqueux \& Christensen(2009{\natexlab{b}})}]{piqueuxchristian2009b}
---. 2009{\natexlab{b}}, Journal of Geophysical Research: Planets, 114

\bibitem[{Piqueux \& Christensen(2011)}]{piqueuxchristian2011}
Piqueux, S., \& Christensen, P.~R. 2011, Journal of Geophysical Research:
  Planets, 116

\bibitem[{Presley \& Christensen(1997{\natexlab{a}})}]{presleychristian1997b}
Presley, M.~A., \& Christensen, P.~R. 1997{\natexlab{a}}, Journal of
  Geophysical Research: Planets, 102, 6535

\bibitem[{Presley \& Christensen(1997{\natexlab{b}})}]{presleychristian1997a}
---. 1997{\natexlab{b}}, Journal of Geophysical Research: Planets, 102, 6535

\bibitem[{Presley \& Christensen(2010)}]{presleychristian2010}
---. 2010, Journal of Geophysical Research: Planets, 115

\bibitem[{Putzig {et~al.}(2014)Putzig, Mellon, Herkenhoff, Phillips, Davis,
  Ewer, \& Bowers}]{putzig2014}
Putzig, N.~E., Mellon, M.~T., Herkenhoff, K.~E., {et~al.} 2014, Icarus, 230,
  64, \dodoi{10.1016/j.icarus.2013.07.010}

\bibitem[{Putzig {et~al.}(2005)Putzig, Mellon, Kretke, \&
  Arvidson}]{putzig2005global}
Putzig, N.~E., Mellon, M.~T., Kretke, K.~A., \& Arvidson, R.~E. 2005, Icarus,
  173, 325

\bibitem[{Rampe {et~al.}(2018)Rampe, Lapotre, Bristow, Arvidson, Morris,
  Achilles, Weitz, Blake, Ming, Morrison, {et~al.}}]{rampe2018sand}
Rampe, E., Lapotre, M., Bristow, T., {et~al.} 2018, Geophysical Research
  Letters, 45, 9488

\bibitem[{Ramsey \& Christensen(1998)}]{ramseychristian1998mineral}
Ramsey, M.~S., \& Christensen, P.~R. 1998, Journal of Geophysical Research:
  Solid Earth, 103, 577

\bibitem[{Richardson {et~al.}(2021)Richardson, Bleacher, Connor, \&
  Glaze}]{richardson2021small}
Richardson, J., Bleacher, J.~E., Connor, C., \& Glaze, L.~S. 2021, Journal of
  Geophysical Research: Planets, e2020JE006620

\bibitem[{Rogers \& Aharonson(2008)}]{rogersahranson2008}
Rogers, A., \& Aharonson, O. 2008, Journal of Geophysical Research: Planets,
  113

\bibitem[{Rogers \& Bandfield(2009)}]{rogersbandfeild2009}
Rogers, A.~D., \& Bandfield, J.~L. 2009, Icarus, 203, 437

\bibitem[{Rogers {et~al.}(2007)Rogers, Bandfield, \&
  Christensen}]{rogersChris2007}
Rogers, A.~D., Bandfield, J.~L., \& Christensen, P.~R. 2007, Journal of
  Geophysical Research: Planets, 112

\bibitem[{Rogers \& Christensen(2007)}]{rogersetal2007}
Rogers, A.~D., \& Christensen, P.~R. 2007, Journal of Geophysical Research:
  Planets, 112

\bibitem[{Rogers \& Fergason(2011)}]{rogersfergonson2011}
Rogers, A.~D., \& Fergason, R.~L. 2011, Journal of Geophysical Research:
  Planets, 116

\bibitem[{Ruff(2004)}]{ruff2004spectral}
Ruff, S.~W. 2004, Icarus, 168, 131

\bibitem[{Sagan {et~al.}(1972)Sagan, Veverka, Fox, Dubisch, Lederberg,
  Levinthal, Quam, Tucker, Pollack, \& Smith}]{sagan1972variable}
Sagan, C., Veverka, J., Fox, P., {et~al.} 1972, Icarus, 17, 346

\bibitem[{Salvatore {et~al.}(2018)Salvatore, Goudge, Bramble, Edwards,
  Bandfield, Amador, Mustard, \& Christensen}]{salvatore2018bulk}
Salvatore, M., Goudge, T., Bramble, M., {et~al.} 2018, Icarus, 301, 76

\bibitem[{Salvatore {et~al.}(2016)Salvatore, Kraft, Edwards, \&
  Christensen}]{salvatore2016geologic}
Salvatore, M., Kraft, M., Edwards, C., \& Christensen, P. 2016, Journal of
  Geophysical Research: Planets, 121, 273

\bibitem[{Salvatore {et~al.}(2014)Salvatore, Mustard, Head~Iii, Rogers, \&
  Cooper}]{salvatore2014dominance}
Salvatore, M., Mustard, J., Head~Iii, J., Rogers, A., \& Cooper, R. 2014, Earth
  and Planetary Science Letters, 404, 261

\bibitem[{Seelos {et~al.}(2014)Seelos, Seelos, Viviano-Beck, Murchie, Arvidson,
  Ehlmann, \& Fraeman}]{seelos2014mineralogy}
Seelos, K.~D., Seelos, F.~P., Viviano-Beck, C.~E., {et~al.} 2014, Geophysical
  Research Letters, 41, 4880

\bibitem[{Sefton-Nash {et~al.}(2014)Sefton-Nash, Teanby, Newman, Clancy, \&
  Richardson}]{sefton2014constraints}
Sefton-Nash, E., Teanby, N., Newman, C., Clancy, R., \& Richardson, M. 2014,
  Icarus, 230, 81

\bibitem[{Silvestro {et~al.}(2020)Silvestro, Chojnacki, Vaz, Cardinale, Yizhaq,
  \& Esposito}]{silvestro2020megaripple}
Silvestro, S., Chojnacki, M., Vaz, D., {et~al.} 2020, Journal of Geophysical
  Research: Planets, 125, e2020JE006446

\bibitem[{Silvestro {et~al.}(2010)Silvestro, Di~Achille, \&
  Ori}]{silvestro2010dune}
Silvestro, S., Di~Achille, G., \& Ori, G. 2010, Geomorphology, 121, 84

\bibitem[{Smith \& Zuber(1998)}]{smithzuber1998}
Smith, D.~E., \& Zuber, M.~T. 1998, Geophysical Research Letters, 25, 4397

\bibitem[{Smith {et~al.}(2000)Smith, Bandfield, \&
  Christensen}]{smith2000separation}
Smith, M.~D., Bandfield, J.~L., \& Christensen, P.~R. 2000, Journal of
  Geophysical Research: Planets, 105, 9589

\bibitem[{Thomas {et~al.}(1981)Thomas, Veverka, Lee, \&
  Bloom}]{thomas1981classification}
Thomas, P., Veverka, J., Lee, S., \& Bloom, A. 1981, Icarus, 45, 124

\bibitem[{Ward {et~al.}(1985)Ward, Doyle, Helm, Weisman, \&
  Witbeck}]{ward1985global}
Ward, A., Doyle, K., Helm, P., Weisman, M., \& Witbeck, N. 1985, Journal of
  Geophysical Research: Solid Earth, 90, 2038

\bibitem[{Williams {et~al.}(2018)Williams, Moersch, \& Fergason}]{williams2018}
Williams, R.~M., Moersch, J.~E., \& Fergason, R.~L. 2018, Earth and Space
  Science, 5, 516

\bibitem[{Wilson \& Zimbelman(2004)}]{wilson2004latitude}
Wilson, S.~A., \& Zimbelman, J.~R. 2004, Journal of Geophysical Research:
  Planets, 109

\bibitem[{Wyatt {et~al.}(2001)Wyatt, Hamilton, McSween~Jr, Christensen, \&
  Taylor}]{wyatt2001analysis}
Wyatt, M.~B., Hamilton, V.~E., McSween~Jr, H.~Y., Christensen, P.~R., \&
  Taylor, L.~A. 2001, Journal of Geophysical Research: Planets, 106, 14711

\end{thebibliography}
\bibliographystyle{aasjournal}

\end{document}